\documentclass[floatfix,a4paper,prd,11pt]{revtex4}
\usepackage{graphics,subfigure}
\usepackage{float}
\usepackage{booktabs}
\usepackage{geometry}
\usepackage{rotating}
\usepackage{yfonts}
\DeclareMathAlphabet{\scr}{U}{rsfs}{m}{n}
\usepackage{supertabular}
\usepackage{amsfonts}
\usepackage{amsmath}
\usepackage{amssymb}
\usepackage[sans]{dsfont}
\usepackage{graphicx}
\usepackage{epsfig}
\usepackage[dvips]{color}
\usepackage{psfrag}
\usepackage{feynarts}
\usepackage{dcolumn}

\newcommand{\newc}{\newcommand}
\newc{\eps}{\epsilon}
\newc{\lam}{\lambda}
\newc{\lamp}{\lambda^{\prime}}
\newc{\Lam}{\Lambda}
\newc{\kap}{\kappa}
\newc{\ra}{\rightarrow}
\newc{\lra}{\leftrightarrow}
\newc{\wtilde}{\widetilde}
\newc{\ie}{{\it i.e.}}
\newc{\eg}{{\it e.g.}}
\newc{\rpv}{\not\!\! M_p}
\newc{\lsim}{\stackrel{<}{\sim}}
\newc{\gsim}{\stackrel{>}{\sim}}
\newc{\beq}{\begin{equation}}
\newc{\eeq}{\end{equation}}
\newc{\beqn}{\begin{eqnarray}}
\newc{\eeqn}{\end{eqnarray}}
\newc{\PLB}{\emph{Phys.Lett.}{\bf{B}}}
\newc{\NPB}{\emph{Nucl.Phys.}{\bf{B}}}
\newc{\mcal}{\mathcal}
\newc{\bsym}{\boldsymbol}
\newc{\nonum}{\nonumber}
\definecolor{Red}{cmyk}{0,1,1,0}
\definecolor{luhn}{rgb}{1,0,0.5}
\definecolor{thor}{rgb}{0,0.6,1}

\newcommand{\dslash}[0]{\ensuremath{{\partial}{\!}{\!}{\!}{\!}{\!}{\:}{\!}{\:}/}}

\newcommand{\prettyfraction}[2]{\ensuremath{{{}^{#1}{\!}/{\!}{}_{#2}}}}
\newcommand{\citeequation}[1]{equation~(\ref{#1})}

\newcommand{\citetable}[1]{Table~\ref{#1}}
\newcommand{\bra}[1]{\ensuremath{\langle {#1} |}}
\newcommand{\ket}[1]{\ensuremath{| {#1} \rangle}}
\newcommand{\dotproduct}[2]{\ensuremath{{#1}{\!}{\!}{\:}{\!}{\:}{\cdot}{\!}{\!}{\:}{\!}{\:}{\!}{\:}{#2}}}
\newcommand{\trace}[1]{\ensuremath{{\mbox{tr}}\left( {#1} \right)}}
\newcommand{\mydefinedby}[0]{\ensuremath{{\:}{\equiv}{\:}}}
\newcommand{\columnspacefixer}[0]{\ensuremath{{\!}{\:}}}

\newcommand{\superpartner}[1]{\ensuremath{{\tilde{{#1}}}}}

\newcommand{\citereference}[1]{reference~\cite{#1}}

\renewcommand{\eqref}[1]{Eq.~(\ref{#1})}
\newc{\eqsref}[2]{Eqs.~(\ref{#1}),\,(\ref{#2})}
\newc{\figref}[1]{Fig.~\ref{#1}}

\newc{\neut}[1]{\tilde{\chi}_{#1}^0}
\newc{\lan}{\mathcal{L}}
\newc{\abs}[1]{\lvert#1\rvert}
\newc{\oas}{\mathcal{O}(\alpha_s^2)}
\newc{\del}{\partial}
\newc{\rb}[1]{\raisebox{1.5ex}[-1.5ex]{#1}}

\def\lsim{\raise0.3ex\hbox{$\;<$\kern-0.75em\raise-1.1ex\hbox{$\sim\;$}}}
\def\gsim{\raise0.3ex\hbox{$\;>$\kern-0.75em\raise-1.1ex\hbox{$\sim\;$}}}
\begin{document}

\title{Bounds on R-parity violating supersymmetric couplings from
  leptonic and semi-leptonic meson decays}

\author{H.~K.~Dreiner}
\affiliation{Physikalisches Institut der Universit\"at Bonn, 
Nu\ss allee 12, 53115 Bonn, Germany}
\author{M.~Kr\"amer}
\affiliation{Institut f\"ur Theoretische Physik E, RWTH Aachen, 
52056 Aachen, Germany}
\author{Ben O'Leary}
\affiliation{School of Physics, University of Edinburgh, Edinburgh EH9 3JZ, Scotland}

\begin{abstract}
  We present a comprehensive update of the bounds on R-Parity
  violating supersymmetric couplings from lepton-flavour- and
  lepton-number-violating decay processes. We consider $\tau$ and
  $\mu$ decays as well as leptonic and semi-leptonic decays of mesons.
  We present several new bounds resulting from $\tau$, $\eta$ and Kaon
  decays and correct some results in the literature concerning
  $B$-meson decays.
\end{abstract}

\maketitle


\section{Introduction}
When extending the symmetries of the Standard Model of particle
physics (SM) \cite{Glashow:1961tr, Weinberg:1967tq} to include
supersymmetry \cite{Wess:1974tw}, the Yukawa couplings are fixed by
the renormalisable superpotential \cite{Sakai:1981pk,Weinberg:1981wj}
\begin{eqnarray}
W &=& \;W_{P_6} + W_{\not{P}_6}^{\not L} + W_{\not{P}_6}^
{\not B} \, , \label{superpot} \\[1mm]
W_{P_6} &=& \epsilon_{ab}  
        \left(h^E_{ij} L_i^a H_d^b {\bar E}_j 
        + h^D_{ij} Q_i^a H_d^b {\bar D}_j 
        + h^U_{ij} Q_i^a H_u^b {\bar U}_j 
        + \mu H_d^aH_u^b \right)\, , \label{superpot-P6} \\[1mm]
W_{\not{P}_6}^{\not L} &=& 
        \epsilon_{ab}\left(\frac{1}{2}\lam_{ijk} L_i^aL_j^b{\bar E}_k 
        + \lam_{ijk}^\prime L_i^aQ_j^b{\bar D}_k 
        + \kap_i L_i^aH_u^b\right) \,, \label{superpot-LV} \\[1mm]
W_{\not{P}_6}^{\not B} &=& \frac{1}{2} \epsilon_{rst}\lam_{ijk}^
{\prime\prime}{\bar U}_i^r{\bar D}_j^s {\bar D}_k^t \,.  \label{superpot-BV}
\end{eqnarray}
Here, $i,j,k=1,2,3$ are generation indices, $a,b=1,2$ are SU(2) and
$r,s,t=1,2,3$ are SU(3) indices. $L,\bar E$ denote the lepton
doublet and singlet left-chiral superfields; $Q,\bar U,\bar D$ denote
the quark doublet and singlet superfields, respectively. $h^E, h^D,
h^U, \lam,\lam',\lam''$ are dimensionless coupling constants and
$\mu,\kap$ are mass mixing parameters.

\medskip

Together the operators in $W_{\not{P}_6}^{\not L}$ and $W_{\not{P}_6}^
{\not B}$ lead to rapid proton decay in disagreement with the
experimental lower bounds on the proton lifetime \cite{
  Shiozawa:1998si}. A possible solution to this problem is to
introduce the discrete $\mathbf{Z}_6$ symmetry, proton hexality, $
\mathbf{P}_6$ \cite{Dreiner:2005rd}, which prohibits both $W_{\not{P}_
  6}^{\not L }$ and $W_{\not{P}_6}^{\not B}$, as well the dangerous
dimension-5 proton decay operators \cite{Sakai:1981pk,
  Dimopoulos:1981dw}; this is the minimal supersymmetric Standard
Model (MSSM) \cite{MSSM_review}. (Note that the widely used discrete
$\mathbf{Z}_2$ symmetry R-parity does not prohibit the dimension-five
proton decay operators.) However, in order to stabilize the proton it
is sufficient to prohibit either the superpotential $W_{\not
  P_6}^{\not L}$ via baryon-triality \cite{Ibanez:1991pr,
  Dreiner:2006xw} or the superpotential $W_{\not P_6}^{\not B}$ via
lepton-parity \cite{ Ibanez:1991pr}.  Lepton-parity is not discrete
gauge anomaly-free \cite{discrete-gauge} and we thus disregard it in
the following.  Baryon-parity has the further advantage of allowing
for non-zero neutrino masses without the need for right-handed
neutrinos. We thus consider here the total superpotential given by
\begin{equation}
W=W_{P_6} + W_{\not{P}_6}^{\not L} \,.
\end{equation}
We shall focus exclusively on the tri-linear couplings. At any given
scale the bi-linear terms $\kap_i L_i H_2$ can be rotated away through
a basis redefinition \cite{Hall:1983id}. This is not true, when
embedding the theory in a more unified model, \eg~supergravity
\cite{Nardi:1996iy,Allanach:2003eb}. However, at $M_{Pl}$ the natural
value is $\kap_i=0$ \cite{Allanach:2003eb}, which leads to $\kap_i\ll
M_W$ at low-energy. Thus the bi-linear terms are mainly relevant for
neutrino masses, see for example \cite{Hirsch:2000ef, Dreiner:2006xw},
and we shall neglect them in the following. The tri-linear operators
in $W_{\not{P}_6}^{\not L}$, lead to novel supersymmetric collider
signatures beyond those of the MSSM \cite{colliders}. In particular,
the operators in $W_{\not{P} _6}^{\not L }$ induce lepton flavour
violation (LFV) as well as lepton number violation, neither of which
has been observed \cite{PDG}.

There is an extensive literature on the resulting bounds on the
operators $W_{\not{P }_6}^{\not L}$ from indirect processes, see
\textit{e.g.}  Refs. \cite{Hall:1983id,Choudhury_Roy,Drei-Pol-Thor,
  Dreiner:2002xg,Agashe:1995qm,Bhattacharyya:1995pr,Saha_Kundu,
  Jang:1997ry, Xu_Wang_Yang, Jang_Kim_Lee,Kim_Ko_Lee,
  Bhattacharyya:1998be, Lebedev:1999vc}, including also several
overviews \cite{Barger_Giudice_Han, Davidson:1993qk, Dreiner:1997uz,
  Bhattacharyya:1997vv, Allanach:1999ic, Herz, Chemtob:2004xr,
  Barbier:2004ez}. However, due to the improved data in particular on
$B$-meson and $\tau$ decays, it is the purpose of this paper to present
a systematic update of the bounds resulting from lepton decays as well
as leptonic and semi-leptonic decays of mesons.  In the process, we
have found several new bounds resulting from $\tau$, $\eta$ and Kaon
decays. We have also found a need to correct some results in the
literature with respect to $B$-meson decays.

Our paper is organized as follows. In
Sect.~\ref{analytic-expressionsa} we start from an effective
Lagrangian, where the supersymmetric scalar fermions have been
integrated out and then present the treatment of the QCD bound state
in Sect.~\ref{analytic-expressionsb}. General analytic expressions for
the decay rates of the various lepton and meson decays are shown in
Sect.~\ref{analytic-expressionsc} .  In Sect.~\ref{numerical-results},
we insert the present experimental results into the analytical
expressions to obtain our new bounds.  These are summarized in Tables
\ref{new_bounds_highlights_table} -
\ref{lambdaprime-lambdaprime_table3}. In Sect.~\ref{discussion}, we
discuss the implications of our results. Formul\ae\ for the meson
decay constants and the general lepton and meson decay matrix elements
are collected in the Appendices.

\section{Theoretical Analysis}
\label{analytic-expressions}
\subsection{Effective Lagrangian}
\label{analytic-expressionsa}

Because the sfermions are constrained to be heavy,
$m_{{\tilde{\nu}},{\tilde{q}}}\; {\gtrsim}\; 100$~GeV~${\gg}\; M_{B}$
(this work does not consider the decays of particles heavier than $B$
mesons), we approximate their propagators as static
$1/m_{\tilde{f}}^2$. This is equivalent to integrating out the
sfermionic degrees of freedom to obtain an effective interaction
Lagrangian \cite{footnote1} and taking only the leading term in an
expansion in inverse sfermion mass, 
\begin{eqnarray}
{\mathcal{L}}_{{\mbox{\scriptsize eff}}} & = & {\sum_{g = 1}^{3}}\; \Bigg\{ {\frac{1}
{{m_{{{\tilde{{\nu}}}_{g}}}^{2}}}} {\lam}_{gab} {\lam}_{gcd}^{{\ast}} 
( {\bar{l}}^{c} P_{R} l^{d} ) ( {\bar{l}}^{b} P_{L} l^{a} ) \label{eq:leffa}  \\
 & & \hspace{2.9mm} + \left[ {\frac{1}{{m_{{{\tilde{{\nu}}}_{g}}}^{2}}}} {\lam}_
{gik} 
{{\lam}'}_{gnm}^{{\ast}} ( {\bar{d}}^{n} P_{R} d^{m} ) ( {\bar{l}}^{k} P_{L} l^{i})
 + {\mbox{ h.c.}} \right] \label{eq:leffb} \\
 & &  \hspace{2.9mm} - \left[ {\frac{1}{{2 m_{{{\tilde{u}}_{g}}}^{2}}}} {{\lam}'}_
{ign} 
{{\lam}'}_{kgm}^{{\ast}} ( {\bar{d}}^{n} {\gamma}^{{\mu}} P_{R} d^{m} ) 
( {\bar{l}}^{k} {\gamma}_{{\mu}} P_{L} l^{i} ) + {\mbox{ h.c.}} \right] 
\label{eq:leffc}
 \\
 & & \hspace{2.9mm} + \left[ {\frac{1}{{2 m_{{{\tilde{d}}_{g}}}^{2}}}} {{\lam}'}_
{img} 
{{\lam}'}_{kng}^{{\ast}} ( {\bar{u}}^{n} {\gamma}^{{\mu}} P_{L} u^{m} ) 
( {\bar{l}}^{k} {\gamma}_{{\mu}} P_{L} l^{i} ) + {\mbox{ h.c.}} \right] \Bigg\}\,, 
\label{eq:leffd}
\label{effective_Lagrangian_equation}
\end{eqnarray}
where (\ref{eq:leffa}) and (\ref{eq:leffb}) result from integrating
out the sneutrino fields, and (\ref{eq:leffc}) and (\ref{eq:leffd})
result from integrating out the up--type and down--type squark fields
respectively, using some Fierz identities.  The index $g$ denotes the
generation. There are additional terms in the effective Lagrangian
which arise when integrating out the charged sleptons, which we do not
consider here: For the product of two $LL\bar E$ or an $LL\bar E$ and
an $LQ\bar D$ operator, these lead to neutrinos in the final state.
Thus lepton flavour violation is not observable in the resulting
lepton or meson decays; for the product of two $LQ\bar D$ operators,
the resulting meson decays are purely hadronic.

\medskip

In the following, we shall assume that the decay is dominated by the
exchange of a sfermion of a single generation, either because it is
lighter than the others or because it has a larger product of
couplings (the 
 double coupling dominance convention). Subsequent
expressions with an index $g$ are thus implicitly for only one value
of $g$, though one may always deduce the general result by replacing
expressions like $|{{\lam}'}_{gjk} {{\lam}'}_{glm} /
m_{{{\tilde{u}}^{g}}}^{2}|^{2}$ with $|{\sum_{g}} {{\lam}'}_{gjk}
{{\lam}'}_{glm} / m_{{{\tilde{u}}^{g}}}^{2}|^{2}$ etc.

It is also assumed in this paper that the sneutrino--higgsino and
squark mixing can be neglected. Such mixings just add to the
notational burden.  If one insists on accounting for mixing, one can
make the replacement $| {{\lam}'}_{gjk} {{\lam}'}_{glm} /
m_{{{\tilde{u}}^{g}}}^{2} |^{2}$ ${\to}$ $| {\sum_{g,x,y}}
{{\lam}'}_{xjk} U_{xg} U_{gy}^{{\dagger}} {{\lam}'}_{ylm} /
m_{{{\tilde{u}}^{g}}}^{2} |^{2}$ etc.\ for squark mixing matrices $U$,
with a similar expression for mixing between the sneutrinos and
Higgses.

\subsection{Meson decay constants}
\label{analytic-expressionsb}
The decay constant, $f_V$, of a vector meson $V$ with momentum $p_V$
is defined as
\begin{equation}
{\bra{0}} {\bar{q}}_{{\alpha}} {\gamma}^{{\mu}} q_{{\beta}} {\ket{{V(
      p_{V} )}}} = H_{V}^{{\alpha}{\beta}} f_{V} m_{V} {\epsilon}_{V}^{{\mu}}\,,
\label{vector_meson_approximation_equationI}
\end{equation}
where ${\epsilon}_{V}^{{\mu}}$ is the polarization vector of $V$,
$m_{V}$ is the vector meson mass, and $H_{V}^{{\alpha}{\beta}}$ is the
coefficient of ${\bar{q}}_{{\alpha}} q_{{\beta}}$ in the quark model
wavefunction of the meson, \eg\ $H_{{{\rho}^{0}}}^{uu} =
{\prettyfraction{1}{{\sqrt{2}}}}$, $H_{{{\rho}^{0}}}^{dd} =
{\prettyfraction{-1}{{\sqrt{2}}}}$.

For a pseudoscalar meson $P$, we use the PCAC condition \cite{pcac}
and define the decay constant $f_{P}$ through the axial vector matrix
element
\begin{equation}
{\bra{0}} {\bar{q}}_{{\alpha}} {\gamma}^{{\mu}} {\gamma}^{5}
q_{{\beta}} {\ket{{P( p_{P} )}}} = i H_{P}^{{\alpha}{\beta}} f_{P} p_{P}^{{\mu}}\,,
\label{PCAC_condition_equationI}
\end{equation}
where $H_{P}^{{\alpha}{\beta}}$ is the analogue of
$H_{V}^{{\alpha}{\beta}}$.  As described in Appendix~\ref{app:qcd},
the equation of motion for the quark fields can be used to derive the
pseudoscalar matrix element from the axial vector matrix element
(\ref{PCAC_condition_equationI}). We find
\begin{equation}
{\bra{0}} {\bar{q}}_{{\alpha}} {\gamma}^{5} q_{{\beta}} {\ket{{P(
      p_{P} )}}} = {\frac{i H_{P}^{{\alpha}{\beta}} f_{P} 
m_{P}^{2}}{{{\mu}_{P}^{{\alpha}{\beta}}}}}\,.
\label{pseudoscalar_meson_approximation_equationI}
\end{equation}
The factor ${\mu}_{P}^{{\alpha}{\beta}}$ is proportional to the sum of
current quark masses $m_\alpha$ and $m_\beta$, \eg\ 
${\mu}_{{{\pi}^{0}}}^{uu} = - 2 m_{u}$ and ${\mu}_{{{\pi}^{0}}}^{dd} =
2 m_{d}$. For the proper definition of ${\mu}_{P}^{{\alpha}{\beta}}$,
a list of the coefficients $H_{P/V}^{{\alpha}{\beta}}$ and more
details see Appendix~\ref{app:qcd}.

\subsection{Decay rates}
\label{analytic-expressionsc}

The Feynman graphs and matrix elements for the various decays
considered in this paper are given in Appendix~\ref{app:me}. Upon
squaring the matrix elements, summing over the final spin states and
averaging over the initial spin states, we arrive at the following
expressions for the decay widths:

\begin{list}{$\bullet$}
{\setlength{\leftmargin}{4mm}}

\item For a heavy lepton $a$ decaying into leptons $b$ and
  $c$ and an anti--lepton $\bar d$,
\begin{equation}
{\Gamma}_{a{\to}bc{\bar{d}}} = {\frac{{m_{{l^{a}}}^5}}{6144 {\pi}^{3} 
m_{{\tilde{{\nu}}}^{g}}^{4}}} ( {\lam}_{gdc}^{2} {\lam}_{gba}^{2} + 
{\lam}_{gcd}^{2} {\lam}_{gab}^{2} + {\lam}_{gdb}^{2} 
{\lam}_{gca}^{2} + {\lam}_{gbd}^{2} {\lam}_{gac}^{2} )\,,
\label{Gamma_a_to_bcd_equation}
\end{equation}
where we approximate the final state (anti--)leptons as massless.

\item[$\bullet$] For a heavy lepton $i$ decaying into a lepton $k$ and
  a vector meson consisting of valence quark $n$ and anti--quark $m$,
  there are two cases: up--type squark--mediated:
\begin{equation}
{\Gamma}_{l^i{\to}l^k+V} = \Bigl| {\sum_{{\mbox{\scriptsize d--type}}}} 
( {{\lam}'}_{ign} {{\lam}'}_{kgm} ) H_{V}^{mn} \Bigr|^{2} 
{\frac{( m_{{l^{i}}}^{2} - m_{V}^{2} )^{2}}{512 {\pi}
    m_{{\tilde{u}}^{g}}^{4}}} {\frac{| f_{V} |^{2} 
( m_{{l^{i}}}^{2} + 2 m_{V}^{2} )}{{m_{{l^{i}}}^{3}}}} 
\left( 1 + {\mathcal{O}}\left( {\frac{{m_{{l^{k}}}}}{{m_{{l^{i}}}}}} \right) \right)
\label{Gamma_to_V_su_equation}
\end{equation}
or down--type squark--mediated:
\begin{equation}
{\Gamma}_{l^i{\to}l^k+V} = \Bigl| {\sum_{{\mbox{\scriptsize u--type}}}} 
( {{\lam}'}_{img} {{\lam}'}_{kng} ) H_{V}^{mn} \Bigr|^{2} 
{\frac{( m_{{l^{i}}}^{2} - m_{V}^{2} )^{2}}{512 {\pi} 
m_{{\tilde{d}}^{g}}^{4}}} {\frac{| f_{V} |^{2} 
( m_{{l^{i}}}^{2} + 2 m_{V}^{2} )}{{m_{{l^{i}}}^{3}}}} 
\left( 1 + {\mathcal{O}}\left( {\frac{{m_{{l^{k}}}}}{{m_{{l^{i}}}}}} \right)\! 
\right)\,,
\label{Gamma_to_V_sd_equation}
\end{equation}
where we have introduced the notation ${\sum_{{\mbox{\scriptsize
d--type}}}}$ to mean only summing over the down--type quarks in the 
meson and ${\sum_ {{\mbox{\scriptsize u--type}}}}$ to mean summing
over the up--type quarks.

\item[$\bullet$] For a heavy lepton $i$ decaying into a lepton $k$ and
  a pseudoscalar meson consisting of valence quark $n$ and anti--quark
  $m$, there are three cases: up--type squark--mediated:
\begin{equation}
{\Gamma}_{l^i{\to}l^k+P} = \Bigl| {\sum_{{\mbox{\scriptsize d--type}}}} 
( {{\lam}'}_{ign} {{\lam}'}_{kgm} ) H_{P}^{mn} \Bigr|^{2} 
{\frac{( m_{{l^{i}}}^{2} - m_{P}^{2} )^{2}}{512 {\pi}
    m_{{\tilde{u}}^{g}}^{4}}} {\frac{| f_{P} |^{2}}{{m_{{l^{i}}}}}} 
\left(\! 1 + {\mathcal{O}}\left( {\frac{{m_{{l^{k}}}}}{{m_{{l^{i}}}}}} \right) 
\right)\,,
\label{Gamma_to_P_su_equation}
\end{equation}
down--type squark--mediated:
\begin{equation}
{\Gamma}_{l^i{\to}l^k+P} = \Bigl| {\sum_{{\mbox{\scriptsize u--type}}}} 
( {{\lam}'}_{img} {{\lam}'}_{kng} ) H_{P}^{mn} \Bigr|^{2} 
{\frac{( m_{{l^{i}}}^{2} - m_{P}^{2} )^{2}}{512 {\pi}
    m_{{\tilde{d}}^{g}}^{4}}} {\frac{| f_{P} |^{2}}{{m_{{l^{i}}}}}} 
\left( 1 + {\mathcal{O}}\left( {\frac{{m_{{l^{k}}}}}{{m_{{l^{i}}}}}} \right) \right)
\label{Gamma_to_P_sd_equation}
\end{equation}
or sneutrino--mediated:
\begin{multline}
{\Gamma}_{l^i{\to}l^k+P}  =  
\left( \Bigl| 
{\sum_{{\mbox{\scriptsize d--type}}}} {\lam}_{gki}^{{\ast}} {{\lam}'}_{gmn} 
{\frac{H_{P}^{mn{\ast}}}{{\mu}_{P}^{mn{\ast}}}} \Bigr|^{2} + 
\Bigl| {\sum_{{\mbox{\scriptsize d--type}}}} {\lam}_{gik} 
{{\lam}'}_{gnm}^{{\ast}} {\frac{H_{P}^{mn{\ast}}}{{\mu}_{P}^{mn{\ast}}}} \Bigr|^{2} \right) \\[2mm]
\times {\frac{( m_{{l^{i}}}^{2} - m_{P}^{2})^{2}}{128
    {\pi} m_{{\tilde{{\nu}}}^{g}}^{4}}} {\frac{| f_{P}|^{2} 
m_{P}^{4}}{m_{{l^{i}}}^{3}}}  
\left( 1 + {\mathcal{O}}\left( {\frac{{m_{{l^{k}}}}}{{m_{{l^{i}}}}}} \right) 
\right)\,.
\label{Gamma_to_P_snu_equation}
\end{multline}
\item[$\bullet$] For a vector meson $V$ decaying into a lepton of
  generation ${k'}$ and an anti--lepton of generation ${i'}$, there
  are again two cases: up--type squark--mediated:
\begin{multline}
{\Gamma}_{V{\to}l^{k'}+\bar{l}^{i'}}  =   \Bigl|
  {\sum_{{\mbox{\scriptsize d--type}}}} 
( {{\lam}'}_{{k'}g{m'}} {{\lam}'}_{{i'}g{n'}} ) H_{V}^{{m'}{n'}}
\Bigr|^{2} {\frac{( m_{V}^{2} - m_{{l^{{i'}}}}^{2} )^{2}}{768 {\pi} 
m_{{\tilde{u}}^{g}}^{4}}} {\frac{| f_{V} |^{2} ( 2 m_{V}^{2} + 
m_{{l^{{i'}}}}^{2} )}{{m_{V}^{3}}}}\\[2mm] 
\times \left( 1 + {\mathcal{O}}\left( 
{\frac{{m_{{l^{{k'}}}}}}{{m_{V}}}} \right) \right)
\label{Gamma_V_to_su_equation}
\end{multline}
or down--type squark--mediated:
\begin{multline}
{\Gamma}_{V{\to}l^{k'}+\bar{l}^{i'}} = \Bigl| {\sum_{{\mbox{\scriptsize u--type}}}} 
( {{\lam}'}_{{k'}{n'}g} {{\lam}'}_{{i'}{m'}g} ) H_{V}^{{m'}{n'}}
\Bigr|^{2} {\frac{( m_{V}^{2} - m_{{l^{{i'}}}}^{2} )^{2}}{768 {\pi} 
m_{{\tilde{d}}^{g}}^{4}}} {\frac{| f_{V} |^{2} ( 2 m_{V}^{2} + 
m_{{l^{{i'}}}}^{2} )}{{m_{V}^{3}}}}\\[2mm] \times \left( 1 + {\mathcal{O}}\left( 
{\frac{{m_{{l^{{k'}}}}}}{{m_{{l^{{i'}}}}}}} \right) \right)\,.
\label{Gamma_V_to_sd_equation}
\end{multline}

\item[$\bullet$] For a pseudoscalar meson $P$ decaying into a lepton
  of generation ${k'}$ and an anti--lepton of generation ${i'}$, there
  are again three cases: up--type squark--mediated:
\begin{equation}
{\Gamma}_{P{\to}l^{k'}+\bar{l}^{i'}} = \Bigl|
  {\sum_{{\mbox{\scriptsize d--type}}}} 
( {{\lam}'}_{{k'}g{m'}} {{\lam}'}_{{i'}g{n'}} ) H_{P}^{{m'}{n'}}
\Bigr|^{2} {\frac{( m_{P}^{2} - m_{{l^{{i'}}}}^{2} )^{2}}{256 {\pi} 
m_{{\tilde{u}}^{g}}^{4}}} {\frac{| f_{P} |^{2}
m_{{l^{{i'}}}}^{2}}{{m_{P}^{3}}}} 
\left( 1 + {\mathcal{O}}\left( {\frac{{m_{{l^{{k'}}}}}}{{m_{P}}}} \right) \right)\,,
\label{Gamma_P_to_su_equation}
\end{equation}
down--type squark--mediated:
\begin{equation}
{\Gamma}_{P{\to}l^{k'}+\bar{l}^{i'}} = \Bigl|
  {\sum_{{\mbox{\scriptsize u--type}}}} 
( {{\lam}'}_{{k'}{n'}g} {{\lam}'}_{{i'}{m'}g} ) H_{P}^{{m'}{n'}}
\Bigr|^{2} {\frac{( m_{P}^{2} - m_{{l^{{i'}}}}^{2} )^{2}}{256 {\pi} 
m_{{\tilde{d}}^{g}}^{4}}} {\frac{| f_{P} |^{2}
m_{{l^{{i'}}}}^{2}}{{m_{P}^{3}}}} 
\left( 1 + {\mathcal{O}}\left( {\frac{{m_{{l^{{k'}}}}}}
{{m_{P}}}} \right) \right)
\label{Gamma_P_to_sd_equation}
\end{equation}
or sneutrino--mediated:
\begin{multline}
{\Gamma}_{P{\to}l^{k'}+\bar{l}^{i'}}  =  
\left( \Bigl| {\sum_{{\mbox{\scriptsize d--type}}}} 
{\lam}_{g{i'}{k'}}^{{\ast}} {{\lam}'}_{g{n'}{m'}}
 {\frac{H_{P}^{{m'}{n'}}}{{\mu}_{P}^{{m'}{n'}{\ast}}}}
\Bigr|^{2} + \Bigl| {\sum_{{\mbox{\scriptsize d--type}}}} {\lam}_{g{k'}{i'}} 
{{\lam}'}_{g{m'}{n'}}^{{\ast}} {\frac{H_{P}^{{m'}{n'}}}{{\mu}_{P}^{{m'}{n'}{\ast}}}} \Bigr|^{2} \right)\\[2mm]
\times {\frac{( m_{P}^{2} - 
m_{{l^{{i'}}}}^{2} )^{2}}{64 {\pi} m_{{\tilde{{\nu}}}^{g}}^{4}}}
| f_{P} |^{2} m_{P}
\left( 1 + {\mathcal{O}}\left( {\frac{{m_{{l^{{k'}}}}}}{{m_{P}}}} \right) \right)\,.
\label{Gamma_P_to_snu_equation}
\end{multline}
\end{list}

\section{Numerical Results}
\label{numerical-results}

We assume for simplicity the \textit{double coupling dominance
  hypothesis}, that the bounds from any one experimental result are
applied to only one product of couplings.

The input values for the various fermion and meson masses and decay
constants are listed in \citetable{input_data_table}.  All the $f_{P}$
values and masses were taken from the 2006 edition of {\em Review of
  Particle Physics} by the Particle Data Group (PDG)~\cite{PDG}.  The
$f_{V}$ values were calculated from $V {\to} e^{+} e^{-}$ according to
\begin{equation}
\Gamma(V\to e^{+} e^{-}) = \frac{4\pi}{3}\frac{\alpha^2}{m_V}f_V^2c_V\,,
\end{equation}
where $c_V$ are factors determined by the electric charge of the
quarks that built up the meson~\cite{C_V_reference}. The experimental
results on lifetimes, decay widths and branching fractions are also
taken from the 2006 review of the PDG~\cite{PDG}.

In Tables \ref{new_bounds_highlights_table},
\ref{very_improved_bounds_highlights_table} and
\ref{combined_better_than_product_bounds_highlights_table} we present
what may be considered the most interesting results of our analysis.
The coupling combinations which had no bounds previously are collected
in \citetable{new_bounds_highlights_table}.  Those combinations which
have improved by a factor of $30$ or more are presented in
\citetable{very_improved_bounds_highlights_table}, and the cases where
the new combined bound is better than the previously published product
of individual bounds are presented in
\citetable{combined_better_than_product_bounds_highlights_table}. Here
and in the following, the symbol $[{\tilde{f}}]$ denotes
$m_{{\tilde{f}}}/{(100~{\mbox{GeV}})} $, \ie~the sfermion mass in
units of $100$~GeV.  This also indicates the mediating sfermion for
the decay. The superscript ${}^{{\dagger}(-)}$ in
\citetable{new_bounds_highlights_table} indicates that this bound
comes from a decay which involves a difference of couplings, so there
could be a cancellation which would lead to the double coupling
dominance hypothesis giving an excessively tight bound. While we also
include very loose bounds in our listings, we note that couplings
$\lam \gsim {\cal O}(2\pi)$ would imply a breakdown of our
perturbative analysis.

\medskip

In Tables~\ref{lambda-lambda_table} to
\ref{lambdaprime-lambdaprime_table6}, we collect all our bounds on the
products of couplings ${\lam}_{ijk}^{(')}{\lam}_{lmn}^{(')}$.  The
results have been arranged so that the number made from reading off
the indices of the couplings to make a six--digit number $ijklmn$
ascends.

In the rightmost columns of Tables~\ref{lambda-lambda_table} to
\ref{lambdaprime-lambdaprime_table6}, ``New'' indicates a previously
unpublished result (see also \citetable{new_bounds_highlights_table}),
``Upd.''~indicates that the bound has been updated and tightened in
this paper, ``Agr.''~indicates that the bound has not changed and we
agree with the previously published result~\cite{footnote2}, and
``Unimp.''~indicates that our bound from decay data is less strong
than the previously published result, which in these cases is from a
different experimental source (\eg~the non--observation of ${\mu}
{\to} e$ in ${}^{48}Ti$ gives a better bound on ${\lam}_{121}
{{\lam}'}_{111} {\tilde{{\nu}}}_{1}^{2}$ than that of ${\pi}^{0} {\to}
e {\bar{{\mu}}}$).  ``Corr.''~indicates that we disagree with the
previously published result~\cite{footnote3},
``Corr.($<$)''~indicating that our result is stronger than the
incorrect previous bound and ``Corr.($>$)''~indicating that our result
is less strong.  The reference in this column gives the previous
published bound.  Where two references are given, the comparison is
between our bound on a product of two couplings and the product of the
bounds on individual couplings.

Note that the $B\to l\bar{l}$ decays can proceed through Standard
Model interactions~\cite{Buras:2001pn}. However, the SM contribution
is suppressed by a small CKM matrix element and by the decay only
arising at one--loop level and has thus been neglected in our
analysis.

\section{Discussion}
\label{discussion}

The bounds presented here generally update those presented in the
literature, with the noted disagreement with some of the bounds coming
from the $B$ meson data.  Many bounds have been improved, some through
tighter experimental decay bounds like those from ${\tau}$ decays,
others through using ${\tau} {\to} K_{S} l^{-}$ instead of ${\tau}
{\to} K^{0} l^{-}$, which also leads to some previously unpublished
bounds.  The ${\eta}$ decay data was also previously unpublished, but
does not seem particularly useful, with bounds of order $10^{2}
{\tilde{f}}^{2}$.  The decay ${\tau} {\to} {\eta} l^{-}$ seems to give
previously unpublished bounds too, which are more stringent. The decay
${\tau} {\to} {\phi} l^{-}$ leads to bounds which are less strong than
those from ${\tau} {\to} {\eta} l^{-}$. Note, however, that ${\tau}
{\to} {\phi} l^{-}$ is free from potential interference effects
induced by the coupling of the mediating squark to both down and
strange quarks.

Assuming that the sfermion masses are of order $100$~GeV and taking
the square root of the bound on a coupling product to be a rough guide
to the bound on each coupling gives an estimate of the couplings
${\lam}$ being of order $0.01$, apart from the very tight bounds from
the non--observation of ${\mu} {\to} e e {\bar{e}}$.  The bounds on
the couplings ${{\lam}'}$ vary considerably, though those involving a
third generation quark are consistently of order $0.01$. Since these
come from $B$ meson decays, they are likely to become even tighter
with more data from $B$ factories.

\begin{table}[p]
\begin{center}
\begin{tabular}{|c|l|l||c|l|}
\hline
Pseudoscalar & Mass     & $f_{P}$         & Fundamental & Mass\\[-2mm]
meson        & (in GeV) & (in GeV)        & fermion     & (in GeV)\\
\hline
${\pi}^{0}$  & $0.135$  & $0.130$         & $e$         & $5.11 {\times} 10^{-4}$\\
\hline
$K_{S}$      & $0.498$  & $0.160$         & ${\mu}$     & $0.106$\\
\hline
$K_{L}$      & $0.498$  & $0.160$         & ${\tau}$    & $1.777$\\
\hline
${\eta}$     & $0.548$  & $0.130$         & $u$         & $3 {\times} 10^{-3}$\\
\hline
${{\eta}'}$  & $0.958$  & $0.172$         & $d$         & $6 {\times} 10^{-3}$\\
\hline
$D_{0}$      & $1.86$   & $0.25$          & $s$         & $0.11$\\
\hline
$B_{d}$      & $5.28$   & $0.2$           & $c$         & $1.25$\\
\hline
$B_{s}$      & $5.37$   & $0.2$           & $b$         & $4.3$\\
\hline 
Vector      & Mass     & $f_{V}$          & \multicolumn{2}{l}{}\\[-2mm]
meson       & (in GeV) & (in GeV)         & \multicolumn{2}{l}{}\\
\cline{1-3}
${\rho}$    & $0.776$  & 0.22             & \multicolumn{2}{l}{}\\
\cline{1-3}
$K^{{ast}}$ & $0.896$  & 0.23             & \multicolumn{2}{l}{}\\
\cline{1-3}
${\phi}$    & $1.020$  & 0.23             & \multicolumn{2}{l}{}\\
\cline{1-3}
$J/{\psi}$  & $3.10$   & 0.41             & \multicolumn{2}{l}{}\\
\cline{1-3}
\end{tabular}
\end{center}
\caption{Input parameters.}
\label{input_data_table}
\end{table}

\begin{table}[p] 
\begin{center}
\begin{tabular}{|c||d@{${\columnspacefixer}$}r|l|}
\hline
Coupling combination & \multicolumn{2}{c|}{Bound}                                  & Decay\\
\hline
${\lambda}_{g21} {{\lambda}'}_{g22}$ & 2.1 & $[ {\superpartner{{\nu}}}_{gL} ]^{2}$ & ${\eta} {\to} {\mu} {\bar{e}} + e {\bar{{\mu}}}$\\
\cline{1-1}
${\lambda}_{g12} {{\lambda}'}_{g22}$ &     &                                       & \\
\hline
${\lambda}_{g13} {{\lambda}'}_{g12}$ & 9.7 & ${\times} 10^{-4}$ $[ {\superpartner{{\nu}}}_{gL} ]^{2}$ & ${\tau} {\to} e K_{S}$\\
\cline{1-1}
${\lambda}_{g31} {{\lambda}'}_{g21}$ &     &                                                          & \\
\hline
${\lambda}_{g23} {{\lambda}'}_{g12}$ & 1.0 & ${\times} 10^{-3}$ $[ {\superpartner{{\nu}}}_{gL} ]^{2}$ & ${\tau} {\to} {\mu} K_{S}$\\
\cline{1-1}
${\lambda}_{g32} {{\lambda}'}_{g21}$ &     &                                                          & \\
\hline
${{\lambda}'}_{1g1} {{\lambda}'}_{3g2}$ & 2.3 & ${\times} 10^{-3}$ $[ {\superpartner{u}}_{gL} ]^{2}$ & ${\tau} {\to} e K_{S}$\\
\hline
${{\lambda}'}_{1g2} {{\lambda}'}_{2g2}$ & 1.5 & ${\times} 10^{+2}$ $[ {\superpartner{u}}_{gL} ]^{2}$ & ${\eta} {\to} {\mu} {\bar{e}} + e {\bar{{\mu}}}$\\
\hline
${{\lambda}'}_{1g2} {{\lambda}'}_{3g2}$ ${}^{{\dagger}(-)}$ & 1.2 & ${\times} 10^{-3}$ $[ {\superpartner{u}}_{gL} ]^{2}$ & ${\tau} {\to} e {\eta}$\\
\hline
${{\lambda}'}_{1g2} {{\lambda}'}_{3g2}$ & 3.4 & ${\times} 10^{-3}$ $[ {\superpartner{u}}_{gL} ]^{2}$ & ${\tau} {\to} e {\phi}$\\
\hline
${{\lambda}'}_{2g1} {{\lambda}'}_{3g2}$ & 2.4 & ${\times} 10^{-3}$ $[ {\superpartner{u}}_{gL} ]^{2}$ & ${\tau} {\to} {\mu} K_{S}$\\
\hline
${{\lambda}'}_{2g2} {{\lambda}'}_{3g2}$ ${}^{{\dagger}(-)}$ & 1.6 & ${\times} 10^{-3}$ $[ {\superpartner{u}}_{gL} ]^{2}$ & ${\tau} {\to} {\mu} {\eta}$\\
\hline
${{\lambda}'}_{2g2} {{\lambda}'}_{3g2}$ & 3.4 & ${\times} 10^{-3}$ $[ {\superpartner{u}}_{gL} ]^{2}$ & ${\tau} {\to} {\mu} {\phi}$\\
\hline
\end{tabular}
\end{center}
\caption{Coupling combinations which had no bounds previous to this
  work. The notation is explained in Sect.~\ref{discussion}}
\label{new_bounds_highlights_table}
\end{table}

\begin{table}[p]
\begin{center}
\begin{tabular}{|c||d@{${\columnspacefixer}$}r|l||d@{${\columnspacefixer}$}r|l|r|}
\hline
Coupling                                  & \multicolumn{3}{c||}{From this work} & \multicolumn{4}{c|}{Previously published}\\
\cline{2-8}
combination                               & \multicolumn{2}{c|}{Bound}      & Decay                        & \multicolumn{2}{c|}{Bound} & Decay & Key\\
\hline
${\lambda}_{g13}$ ${{\lambda}'}_{g22}$    & 4.6 & ${\times} 10^{-4}$ $[ {\superpartner{{\nu}}}_{gL} ]^{2}$ & ${\tau} {\to} e {\eta}$ & 1.6 & ${\times} 10^{-2}$ $[ {\superpartner{{\nu}}}_{gL} ]^{2}$ & ${\tau} {\to} e {\eta}$ & Upd.~\cite{RPV_review382}\\
\cline{1-1}
${\lambda}_{g31}$ ${{\lambda}'}_{g22}$    & & & & & & & \\
\hline
${\lambda}_{g13}$ ${{\lambda}'}_{g21}$    & 9.7 & ${\times} 10^{-4}$ $[ {\superpartner{{\nu}}}_{gL} ]^{2}$ & ${\tau} {\to} e K_{S}$ & 8.5 & ${\times} 10^{-2}$ $[ {\superpartner{{\nu}}}_{gL} ]^{2}$ & ${\tau} {\to} e K^{0}$ & Upd.~\cite{RPV_review382}\\
\cline{1-1}
${\lambda}_{g31}$ ${{\lambda}'}_{g12}$    & & & & & & & \\
\hline
${\lambda}_{g23}$ ${{\lambda}'}_{g21}$    & 1.0 & ${\times} 10^{-3}$ $[ {\superpartner{{\nu}}}_{gL} ]^{2}$ & ${\tau} {\to} {\mu} K_{S}$ & 7.6 & ${\times} 10^{-2}$ $[ {\superpartner{{\nu}}}_{gL} ]^{2}$ & ${\tau} {\to} {\mu} K^{0}$ & Upd.~\cite{RPV_review382}\\
\cline{1-1}
${\lambda}_{g32}$ ${{\lambda}'}_{g12}$    & & & & & & & \\
\hline
\end{tabular}
\end{center}
\caption{Coupling combinations which have improved by a factor of $30$ or more compared to those published before this work.}
\label{very_improved_bounds_highlights_table}
\end{table}

\begin{table}[p]
\begin{center}
\begin{tabular}{|c||d@{${\columnspacefixer}$}r|l||d@{${\columnspacefixer}$}r|l|r|}
\hline
Coupling                                  & \multicolumn{3}{c||}{From this work}                        & \multicolumn{4}{c|}{Previously published}\\
\cline{2-8}
combination                               & \multicolumn{2}{c|}{Bound}                                     & Decay & \multicolumn{2}{c|}{Bound} & Decay & Key\\
\hline
${{\lambda}'}_{11g}$ ${{\lambda}'}_{31g}$ & 1.2 & ${\times} 10^{-3}$ $[ {\superpartner{d}}_{Rg} ]^{2}$     & ${\tau} {\to} e {\pi}^{0}$    & 0.02          & $[ {\superpartner{d}}_{Rg} ]$ & APV in Cs & Upd.~\cite{Barbier:2004ez},\\
            &     &                                                      &                                 & {\times} 0.12 & $[ {\superpartner{d}}_{Rg} ]$ & ${\frac{{\tau} {\to} {\pi}^{-} {\nu}}{{{\pi}^{-} {\to} {\mu} {\bar{{\nu}}}}}}$ & \cite{Barbier:2004ez}\\
\hline
${{\lambda}'}_{12g}$ ${{\lambda}'}_{21g}$ & 9.0 & ${\times} 10^{-3}$ $[ {\superpartner{d}}_{Rg} ]^{2}$     & $D^{0} {\to} {\mu} {\bar{e}}$ & 0.21          & $[ {\superpartner{d}}_{Rg} ]$                   & $A_{FB}^{c}$ & Upd.~\cite{Barbier:2004ez},\\
            &     &                                                      &                                 & {\times} 5.9 & ${\times}10^{-2}$ $[ {\superpartner{d}}_{Rg} ]$ & ${\frac{{{\pi}^{-} {\to} e {\bar{{\nu}}}}}{{{\pi}^{-} {\to} {\mu} {\bar{{\nu}}}}}}$ & \cite{Herz}\\
\hline
\end{tabular}
\end{center}
\caption{Coupling combinations where the combined bound is now better than the product of the individual bounds.}
\label{combined_better_than_product_bounds_highlights_table}
\end{table}

\begin{table}[p]
\begin{center}
\begin{tabular}{|c|d@{${\columnspacefixer}$}r|l||d@{${\columnspacefixer}$}r|r|}
\hline
$( {\lambda}_{ijk}$ ${\lambda}_{lmn} )$ & \multicolumn{3}{c||}{From this work}                & \multicolumn{3}{c|}{Previously published}\\
\cline{2-7}
$ijk$ $lmn$ & \multicolumn{2}{c|}{Bound}                                     & Decay                                    & \multicolumn{2}{c|}{Bound}   & Key\\
\hline
$121$ $123$ & 7.0 & ${\times} 10^{-4}$ $[ {\superpartner{{\nu}}}_{L1} ]^{2}$ & ${\tau} {\to} {\mu} e {\bar{{\mu}}}$     & 2.1 & ${\times} 10^{-3}$ $[ {\superpartner{{\nu}}}_{L1} ]^{2}$ & Upd.~\cite{RPV_review359}\\
\hline
$121$ $131$ & 6.8 & ${\times} 10^{-4}$ $[ {\superpartner{{\nu}}}_{L1} ]^{2}$ & ${\tau} {\to} {\mu} e {\bar{e}}$         & 2.0 & ${\times} 10^{-3}$ $[ {\superpartner{{\nu}}}_{L1} ]^{2}$ & Upd.~\cite{RPV_review359}\\
\hline
$121$ $132$ & 5.6 & ${\times} 10^{-4}$ $[ {\superpartner{{\nu}}}_{L1} ]^{2}$ & ${\tau} {\to} {\mu} {\mu} {\bar{e}}$     & 1.9 & ${\times} 10^{-3}$ $[ {\superpartner{{\nu}}}_{L1} ]^{2}$ & Upd.~\cite{RPV_review359}\\
\hline
$122$ $123$ & 6.8 & ${\times} 10^{-4}$ $[ {\superpartner{{\nu}}}_{L1} ]^{2}$ & ${\tau} {\to} {\mu} {\mu} {\bar{{\mu}}}$ & 2.2 & ${\times} 10^{-3}$ $[ {\superpartner{{\nu}}}_{L1} ]^{2}$ & Upd.~\cite{RPV_review359}\\
\hline
$122$ $131$ & 7.0 & ${\times} 10^{-4}$ $[ {\superpartner{{\nu}}}_{L1} ]^{2}$ & ${\tau} {\to} {\mu} e {\bar{{\mu}}}$     & 2.1 & ${\times} 10^{-3}$ $[ {\superpartner{{\nu}}}_{L1} ]^{2}$ & Upd.~\cite{RPV_review359}\\
\hline
$122$ $132$ & 6.8 & ${\times} 10^{-4}$ $[ {\superpartner{{\nu}}}_{L1} ]^{2}$ & ${\tau} {\to} {\mu} {\mu} {\bar{{\mu}}}$ & 2.2 & ${\times} 10^{-3}$ $[ {\superpartner{{\nu}}}_{L1} ]^{2}$ & Upd.~\cite{RPV_review359}\\
\hline
\end{tabular}
\end{center}
\caption{Bounds on $\mathbf{({\lambda}_{ijk} {\lambda}_{lmn})}$: all but those from ${\mu} {\to} e e {\bar{e}}$ are updated from \citereference{RPV_review359}.  The presented ${\mu} {\to} e e {\bar{e}}$ bounds agree with those in \citereference{Choudhury_Roy}.}
\label{lambda-lambda_table}
\end{table}

\begin{table}[p]
\begin{center}
\begin{tabular}{|c||d@{${\columnspacefixer}$}r|l||d@{${\columnspacefixer}$}r|l|r|}
\hline
$( {\lambda}_{ijk}$ ${{\lambda}'}_{lmn} )$ & \multicolumn{3}{c||}{From this work}      & \multicolumn{4}{c|}{Previously published}\\
\cline{2-8}
$ijk$ $lmn$                                & \multicolumn{2}{c|}{Bound}      & Decay                             & \multicolumn{2}{c|}{Bound} & Decay & Key\\
\hline
$121$ $111$ & 1.2  & ${\times} 10^{-2}$ $[ {\superpartner{{\nu}}}_{L1} ]^{2}$ & ${\pi}^{0} {\to} e {\bar{{\mu}}}$ & 2.1 & ${\times} 10^{-8}$ $[ {\superpartner{{\nu}}}_{L1} ]^{2}$ & ${\mu} {\to} e$ in ${}^{48}$Ti & Unimp.~\cite{RPV_review382-383}\\
\cline{2-4}
            & 0.39  & $[ {\superpartner{{\nu}}}_{L1} ]^{2}$ & ${\eta} {\to} {\mu} {\bar{e}} + e {\bar{{\mu}}}$ & & & & \\
\cline{2-4}
            & 0.41 & $[ {\superpartner{{\nu}}}_{L1} ]^{2}$ & ${\pi}^{0} {\to} {\mu} {\bar{e}}$ & & & & \\
\cline{2-4}
            & 16   & $[ {\superpartner{{\nu}}}_{L1} ]^{2}$ & ${{\eta}'} {\to} {\mu} {\bar{e}} / e {\bar{{\mu}}}$ & & & & \\
\hline
$121$ $112$ & 6.7  & ${\times} 10^{-9}$ $[ {\superpartner{{\nu}}}_{L1} ]^{2}$ & $K_{L}^{0} {\to} {\mu} {\bar{e}} / e {\bar{{\mu}}}$ & 6 & ${\times} 10^{-9}$ $[ {\superpartner{{\nu}}}_{L1} ]^{2}$ & $K_{L}^{0} {\to} {\mu} {\bar{e}} / e {\bar{{\mu}}}$ & ${}^{{\dagger}(-)}$ Agr.~\cite{Drei-Pol-Thor}\\
\hline
$121$ $113$ & 1.3  & ${\times} 10^{-5}$ $[ {\superpartner{{\nu}}}_{L1} ]^{2}$ & $B_{d}^{0} {\to} {\mu} {\bar{e}}$ & 2.3 & ${\times} 10^{-5}$ $[ {\superpartner{{\nu}}}_{L1} ]^{2}$ & $B_{d}^{0} {\to} {\mu} {\bar{e}}$ & Corr.($<$)~\cite{Saha_Kundu}\\
\hline
\end{tabular}
\end{center}
\caption{Bounds on $\mathbf{({\lambda}_{ijk} {{\lambda}'}_{lmn})}$.}
\label{lambda-lambdaprime_table1}
\end{table}

\begin{table}[p]
\begin{center}
\begin{tabular}{|c||d@{${\columnspacefixer}$}r|l||d@{${\columnspacefixer}$}r|l|r|}
\hline
$( {\lambda}_{ijk}$ ${{\lambda}'}_{lmn} )$ & \multicolumn{3}{c||}{From this work}  & \multicolumn{4}{c|}{Previously published}\\
\cline{2-8}
$ijk$ $lmn$                                & \multicolumn{2}{c|}{Bound}      & Decay                         & \multicolumn{2}{c|}{Bound} & Decay & Key\\
\hline
$132$ $111$ & 6.7  & ${\times} 10^{-5}$ $[ {\superpartner{{\nu}}}_{L1} ]^{2}$ & ${\tau} {\to} {\mu} {\eta}$ & 1.7 & ${\times} 10^{-3}$ $[ {\superpartner{{\nu}}}_{L1} ]^{2}$ & ${\tau} {\to} {\mu} {\eta}$ & Upd.~\cite{RPV_review382}\\
\cline{2-4}
            & 1.0  & ${\times} 10^{-3}$ $[ {\superpartner{{\nu}}}_{L1} ]^{2}$ & ${\tau} {\to} {\mu} {\pi}^{0}$ & & & & \\
\hline
$132$ $112$ & 1.0  & ${\times} 10^{-3}$ $[ {\superpartner{{\nu}}}_{L1} ]^{2}$ & ${\tau} {\to} {\mu} K_{S}$ & 7.6 & ${\times} 10^{-2}$ $[ {\superpartner{{\nu}}}_{L1} ]^{2}$ & ${\tau} {\to} {\mu} K^{0}$ & Upd.~\cite{RPV_review382}\\
\hline
$132$ $113$ & 2.2  & ${\times} 10^{-4}$ $[ {\superpartner{{\nu}}}_{L1} ]^{2}$ & $B_{d}^{0} {\to} {\tau} {\bar{{\mu}}}$ & 6.2 & ${\times} 10^{-4}$ $[ {\superpartner{{\nu}}}_{L1} ]^{2}$ & $B_{d}^{0} {\to} {\mu} {\bar{{\tau}}}$ & Corr.($<$)~\cite{Saha_Kundu}\\
\hline
$132$ $121$ & 1.0  & ${\times} 10^{-3}$ $[ {\superpartner{{\nu}}}_{L1} ]^{2}$ & ${\tau} {\to} {\mu} K_{S}$ & & none & n/a & New\\
\hline
$132$ $122$ & 3.7  & ${\times} 10^{-4}$ $[ {\superpartner{{\nu}}}_{L1} ]^{2}$ & ${\tau} {\to} {\mu} {\eta}$ & 1.7 & ${\times} 10^{-2}$ $[ {\superpartner{{\nu}}}_{L1} ]^{2}$ & ${\tau} {\to} {\mu} {\eta}$ & Upd.~\cite{RPV_review382}\\
\hline
\end{tabular}
\end{center}
\caption{Bounds on $\mathbf{({\lambda}_{ijk} {{\lambda}'}_{lmn})}$ continued.}
\label{lambda-lambdaprime_table2}
\end{table}

\begin{table}[p]
\begin{center}
\begin{tabular}{|c||d@{${\columnspacefixer}$}r|l||d@{${\columnspacefixer}$}r|l|r|}
\hline
$( {\lambda}_{ijk}$ ${{\lambda}'}_{lmn} )$ & \multicolumn{3}{c||}{From this work}  & \multicolumn{4}{c|}{Previously published}\\
\cline{2-8}
$ijk$ $lmn$                                & \multicolumn{2}{c|}{Bound}       & Decay                         & \multicolumn{2}{c|}{Bound} & Decay & Key\\
\hline
$231$ $211$ & 8.5  & ${\times} 10^{-5}$ $[ {\superpartner{{\nu}}}_{L2} ]^{2}$ & ${\tau} {\to} e {\eta}$ & 1.6 & ${\times} 10^{-3}$ $[ {\superpartner{{\nu}}}_{L2} ]^{2}$ & ${\tau} {\to} e {\eta}$ & Upd.~\cite{RPV_review382}\\
\cline{2-4}
            & 7.1  & ${\times} 10^{-4}$ $[ {\superpartner{{\nu}}}_{L2} ]^{2}$ & ${\tau} {\to} e {\pi}^{0}$ & & & & \\
\hline
$231$ $212$ & 9.7  & ${\times} 10^{-4}$ $[ {\superpartner{{\nu}}}_{L2} ]^{2}$ & ${\tau} {\to} e K_{S}$ & 8.5 & ${\times} 10^{-2}$ $[ {\superpartner{{\nu}}}_{L2} ]^{2}$ & ${\tau} {\to} e K^{0}$ & Upd.~\cite{RPV_review382}\\
\hline
$231$ $213$ & 3.7  & ${\times} 10^{-4}$ $[ {\superpartner{{\nu}}}_{L2} ]^{2}$ & $B_{d}^{0} {\to} {\tau} {\bar{e}}$ & 4.9 & ${\times} 10^{-4}$ $[ {\superpartner{{\nu}}}_{L2} ]^{2}$ & $B_{d}^{0} {\to} {\tau} {\bar{e}}$ & Corr.($<$)~\cite{Saha_Kundu}\\
\hline
$231$ $221$ & 9.7  & ${\times} 10^{-4}$ $[ {\superpartner{{\nu}}}_{L2} ]^{2}$ & ${\tau} {\to} e K_{S}$ & & none & n/a & New\\
\hline
$231$ $222$ & 4.6  & ${\times} 10^{-4}$ $[ {\superpartner{{\nu}}}_{L2} ]^{2}$ & ${\tau} {\to} e {\eta}$ & 1.6 & ${\times} 10^{-2}$ $[ {\superpartner{{\nu}}}_{L2} ]^{2}$ & ${\tau} {\to} e {\eta}$ & Upd.~\cite{RPV_review382}\\
\hline
\end{tabular}
\end{center}
\caption{Bounds on $\mathbf{({\lambda}_{ijk} {{\lambda}'}_{lmn})}$ continued.}
\label{lambda-lambdaprime_table3}
\end{table}

\begin{table}[p]
\begin{center}
\begin{tabular}{|c||d@{${\columnspacefixer}$}r|l||d@{${\columnspacefixer}$}r|l|r|}
\hline
$( {\lambda}_{ijk}$ ${{\lambda}'}_{lmn} )$ & \multicolumn{3}{c||}{From this work}  & \multicolumn{4}{c|}{Previously published}\\
\cline{2-8}
$ijk$ $lmn$                                & \multicolumn{2}{c|}{Bound}       & Decay                         & \multicolumn{2}{c|}{Bound} & Decay & Key\\
\hline
$313$ $311$ & 8.5  & ${\times} 10^{-5}$ $[ {\superpartner{{\nu}}}_{L3} ]^{2}$ & ${\tau} {\to} e {\eta}$ & 1.6 & ${\times} 10^{-3}$ $[ {\superpartner{{\nu}}}_{L1} ]^{2}$ & ${\tau} {\to} e {\eta}$ & Upd.~\cite{RPV_review382}\\
\cline{2-4}
            & 7.1  & ${\times} 10^{-4}$ $[ {\superpartner{{\nu}}}_{L3} ]^{2}$ & ${\tau} {\to} e {\pi}^{0}$ & & & & \\
\hline
$313$ $312$ & 9.7  & ${\times} 10^{-4}$ $[ {\superpartner{{\nu}}}_{L3} ]^{2}$ & ${\tau} {\to} e K_{S}$ & & none & n/a & New\\
\hline
$313$ $313$ & 3.7  & ${\times} 10^{-4}$ $[ {\superpartner{{\nu}}}_{L3} ]^{2}$ & $B_{d}^{0} {\to} e {\bar{{\tau}}}$ & 4.9 & ${\times} 10^{-4}$ $[ {\superpartner{{\nu}}}_{L3} ]^{2}$ & $B_{d}^{0} {\to} e {\bar{{\tau}}}$ & Corr.($<$)~\cite{Saha_Kundu}\\
\hline
$313$ $321$ & 9.7  & ${\times} 10^{-4}$ $[ {\superpartner{{\nu}}}_{L3} ]^{2}$ & ${\tau} {\to} e K_{S}$ & 8.5 & ${\times} 10^{-2}$ $[ {\superpartner{{\nu}}}_{L3} ]^{2}$ & ${\tau} {\to} e K^{0}$ & Upd.~\cite{RPV_review382}\\
\hline
$313$ $322$ & 4.6  & ${\times} 10^{-4}$ $[ {\superpartner{{\nu}}}_{L3} ]^{2}$ & ${\tau} {\to} e {\eta}$ & 1.6 & ${\times} 10^{-2}$ $[ {\superpartner{{\nu}}}_{L3} ]^{2}$ & ${\tau} {\to} e {\eta}$ & Upd.~\cite{RPV_review382}\\
\hline
\end{tabular}
\end{center}
\caption{Bounds on $\mathbf{({\lambda}_{ijk} {{\lambda}'}_{lmn})}$ continued.}
\label{lambda-lambdaprime_table4}
\end{table}

\begin{table}[p]
\begin{center}
\begin{tabular}{|c||d@{${\columnspacefixer}$}r|l||d@{${\columnspacefixer}$}r|l|r|}
\hline
$( {{\lambda}'}_{ijk}$ ${{\lambda}'}_{lmn} )$ & \multicolumn{3}{c||}{From this work} & \multicolumn{4}{c|}{Previously published}\\
\cline{2-8}
$ijk$ $lmn$ & \multicolumn{2}{c|}{Bound}               & Decay                         & \multicolumn{2}{c|}{Bound}                                       & Decay & Key\\
\hline
$111$ $113$ & 2.6  & ${\times} 10^{-2}$ $[ {\superpartner{u}}_{L1} ]^{2}$ & $B_{d}^{0} {\to} e {\bar{e}}$ & 0.03          & $[ {\superpartner{u}}_{L1} ]$ & APV in Cs & Unimp.~\cite{Barbier:2004ez},\\
            &      &                                                      &                               & {\times} 0.18 & $[ {\superpartner{u}}_{L1} ]$ & $A_{FB}^{b}$ & \cite{Barbier:2004ez}\\
\hline
$111$ $211$ & 0.36 & $[ {\superpartner{d}}_{R1} ]^{2}$ & ${\pi}^{0} {\to} e {\bar{{\mu}}}$ & 4.5 & ${\times} 10^{-8}$ $[ {\superpartner{d}}_{R1} ]^{2}$ & ${\mu} {\to} e$ in ${}^{48}$Ti & Unimp.~\cite{RPV_review382-383}\\
\cline{2-4}
            & 11   & $[ {\superpartner{d}}_{R1} ]^{2}$ & ${\pi}^{0} {\to} {\mu} {\bar{e}}$ & & & & \\
\cline{2-4}
            & 1.5  & ${\times} 10^{+2}$ $[ {\superpartner{d}}_{R1} ]^{2}$ & ${\eta} {\to} {\mu} {\bar{e}} + e {\bar{{\mu}}}$ & & & & \\
\cline{2-4}
            & 1.9  & ${\times} 10^{+4}$ $[ {\superpartner{d}}_{R1} ]^{2}$ & ${{\eta}'} {\to} {\mu} {\bar{e}} / e {\bar{{\mu}}}$ & & & & \\
\hline
\end{tabular}
\end{center}
\caption{Bounds on $\mathbf{({{\lambda}'}_{ijk} {{\lambda}'}_{lmn})}$.}
\label{lambdaprime-lambdaprime_table1}
\end{table}

\begin{table}[p]
\begin{center}
\begin{tabular}{|c||d@{${\columnspacefixer}$}r|l||d@{${\columnspacefixer}$}r|l|r|}
\hline
$( {{\lambda}'}_{ijk}$ ${{\lambda}'}_{lmn} )$ & \multicolumn{3}{c||}{From this work} & \multicolumn{4}{c|}{Previously published}\\
\cline{2-8}
$ijk$ $lmn$ & \multicolumn{2}{c|}{Bound}                                 & Decay                               & \multicolumn{2}{c|}{Bound}                    & Decay & Key\\
\hline
$112$ $212$ & 76   & $[ {\superpartner{u}}_{L1} ]^{2}$ & ${\eta} {\to} {\mu} {\bar{e}} + e {\bar{{\mu}}}$ & & none & n/a & New\\
\cline{2-4}
            & 1.1  & ${\times} 10^{+5}$ $[ {\superpartner{u}}_{L1} ]^{2}$ & ${{\eta}'} {\to} {\mu} {\bar{e}}$ & & & & \\
\hline
$112$ $213$ & 9.4  & ${\times} 10^{-3}$ $[ {\superpartner{u}}_{L1} ]^{2}$ & $B_{s}^{0} {\to} e {\bar{{\mu}}}$ & 2.7 & ${\times} 10^{-4}$ $[ {\superpartner{u}}_{L1} ]^{2}$ & $b {\to} s e {\bar{{\mu}}}$ & Unimp.~\cite{Jang:1997ry}\\
\hline
$112$ $222$ & 2.8  & ${\times} 10^{-2}$ $[ {\superpartner{d}}_{R2} ]^{2}$ & $D^{0} {\to} e {\bar{{\mu}}}$ & 0.02          & $[ {\superpartner{d}}_{R2} ]$ & APV in Cs & Unimp.~\cite{Barbier:2004ez},\\
            &      &                                                      &                               & {\times} 0.21 & $[ {\superpartner{d}}_{R2} ]$ & ${\frac{{\tau} {\to} {\pi}^{-} {\nu}}{{{\pi}^{-} {\to} {\mu} {\bar{{\nu}}}}}}$ & \cite{RPV_review277}\\
\hline
$112$ $311$ & 2.3  & ${\times} 10^{-3}$ $[ {\superpartner{u}}_{L1} ]^{2}$ & ${\tau} {\to} e K_{S}$ & 2.7 & ${\times} 10^{-3}$ $[ {\superpartner{u}}_{L1} ]^{2}$ & ${\tau} {\to} e K^{0{\ast}}$ & Upd.~\cite{RPV_review382}\\
\hline
\end{tabular}
\end{center}
\caption{Bounds on $\mathbf{({{\lambda}'}_{ijk} {{\lambda}'}_{lmn})}$ continued.}
\label{lambdaprime-lambdaprime_table2}
\end{table}

\begin{table}[p]
\begin{center}
\begin{tabular}{|c||d@{${\columnspacefixer}$}r|l||d@{${\columnspacefixer}$}r|l|r|}
\hline
$( {{\lambda}'}_{ijk}$ ${{\lambda}'}_{lmn} )$ & \multicolumn{3}{c||}{From this work} & \multicolumn{4}{c|}{Previously published}\\
\cline{2-8}
$ijk$ $lmn$ & \multicolumn{2}{c|}{Bound}                                 & Decay                               & \multicolumn{2}{c|}{Bound}                     & Decay & Key\\
\hline
$121$ $123$ & 2.6  & ${\times} 10^{-2}$ $[ {\superpartner{u}}_{L2} ]^{2}$ & $B_{d}^{0} {\to} e {\bar{e}}$ & 0.03          & $[ {\superpartner{u}}_{L2} ]$ & APV in Cs    & Unimp.~\cite{Barbier:2004ez},\\
            &      &                                                      &                               & {\times} 0.18 & $[ {\superpartner{u}}_{L2} ]$ & $A_{FB}^{b}$ & \cite{Barbier:2004ez}\\
\hline
$121$ $211$ & 9.0  & ${\times} 10^{-3}$ $[ {\superpartner{d}}_{R1} ]^{2}$ & $D^{0} {\to} {\mu} {\bar{e}}$ & 0.21         & $[ {\superpartner{d}}_{R1} ]$                   & $A_{FB}^{c}$ & Upd.~\cite{Barbier:2004ez},\\
            &      &                                                      &                               & {\times} 5.9 & ${\times}10^{-2}$ $[ {\superpartner{d}}_{R1} ]$ & ${\frac{{{\pi}^{-} {\to} e {\bar{{\nu}}}}}{{{\pi}^{-} {\to} {\mu} {\bar{{\nu}}}}}}$ & \cite{Herz}\\
\hline
$121$ $221$ & 1.6  & $[ {\superpartner{d}}_{R1} ]^{2}$ & $J/{\psi} {\to} {\mu} {\bar{e}} / e {\bar{{\mu}}}$ & 0.21          & $[ {\superpartner{d}}_{R1} ]$ & $A_{FB}^{c}$ & Unimp.~\cite{Barbier:2004ez},\\
            &      &                                   &                                                    & {\times} 0.21 & $[ {\superpartner{d}}_{R1} ]$ & ${\frac{{D^{0} {\to} {\nu} {\bar{{\mu}}} K^{-}}}{{D^{0} {\to} {\nu} {\bar{e}} K^{-}}}}$ & \cite{RPV_review277}\\
\hline
\end{tabular}
\end{center}
\caption{Bounds on $\mathbf{({{\lambda}'}_{ijk} {{\lambda}'}_{lmn})}$ continued.}
\label{lambdaprime-lambdaprime_table3}
\end{table}

\begin{table}[p]
\begin{center}
\begin{tabular}{|c||d@{${\columnspacefixer}$}r|l||d@{${\columnspacefixer}$}r|l|r|}
\hline
$( {{\lambda}'}_{ijk}$ ${{\lambda}'}_{lmn} )$ & \multicolumn{3}{c||}{From this work} & \multicolumn{4}{c|}{Previously published}\\
\cline{2-8}
$ijk$ $lmn$ & \multicolumn{2}{c|}{Bound}                                 & Decay                               & \multicolumn{2}{c|}{Bound}                     & Decay & Key\\
\hline
$122$ $223$ & 9.4  & ${\times} 10^{-3}$ $[ {\superpartner{u}}_{L2} ]^{2}$ & $B_{s}^{0} {\to} e {\bar{{\mu}}}$ & 2.7 & ${\times} 10^{-4}$ $[ {\superpartner{u}}_{L2} ]^{2}$ & $b {\to} s e {\bar{{\mu}}}$ & Unimp.~\cite{Jang:1997ry}\\
\hline
$122$ $321$ & 2.3  & ${\times} 10^{-3}$ $[ {\superpartner{u}}_{L2} ]^{2}$ & ${\tau} {\to} e K_{S}$ & 2.7 & ${\times} 10^{-3}$ $[ {\superpartner{u}}_{L2} ]^{2}$ & ${\tau} {\to} e K^{0{\ast}}$ & Upd.~\cite{RPV_review382}\\
\cline{2-4}
            & 2.9  & ${\times} 10^{-3}$ $[ {\superpartner{u}}_{L2} ]^{2}$ & ${\tau} {\to} e K^{{\ast}0}$ & & & & \\
\hline
$122$ $322$ & 5.9  & $[ {\superpartner{d}}_{R2} ]^{2}$ & $J/{\psi} {\to} {\tau} {\bar{e}} / e {\bar{{\tau}}}$ & 0.21          & $[ {\superpartner{d}}_{R2} ]$ & $A_{FB}^{c}$ & Unimp.~\cite{Barbier:2004ez},\\
            &      &                                   &                                                      & {\times} 0.52 & $[ {\superpartner{d}}_{R2} ]$ & ${\frac{{D_{s}^{-} {\to} {\tau} {\bar{{\nu}}}}}{{D_{s}^{-} {\to} {\mu} {\bar{{\nu}}}}}}$ & \cite{RPV_review271}\\
\hline
$122$ $322$ & 1.2  & ${\times} 10^{-3}$ $[ {\superpartner{u}}_{L2} ]^{2}$ & ${\tau} {\to} e {\eta}$ & & none & n/a & ${}^{{\dagger}(-)}$ New\\
\hline
\end{tabular}
\end{center}
\caption{Bounds on $\mathbf{({{\lambda}'}_{ijk} {{\lambda}'}_{lmn})}$ continued.}
\label{lambdaprime-lambdaprime_table4}
\end{table}

\begin{table}[p]
\begin{center}
\begin{tabular}{|c||d@{${\columnspacefixer}$}r|l||d@{${\columnspacefixer}$}r|l|r|}
\hline
$( {{\lambda}'}_{ijk}$ ${{\lambda}'}_{lmn} )$ & \multicolumn{3}{c||}{From this work} & \multicolumn{4}{c|}{Previously published}\\
\cline{2-8}
$ijk$ $lmn$ & \multicolumn{2}{c|}{Bound}                                 & Decay                               & \multicolumn{2}{c|}{Bound}                     & Decay & Key\\
\hline
$132$ $231$ & 2.7  & ${\times} 10^{-7}$ $[ {\superpartner{u}}_{L3} ]^{2}$ & $K_{L}^{0} {\to} {\mu} {\bar{e}} / e {\bar{{\mu}}}$ & 3 & ${\times} 10^{-7}$ $[ {\superpartner{u}}_{L3} ]^{2}$ & $K_{L}^{0} {\to} {\mu} {\bar{e}} / e {\bar{{\mu}}}$ & ${}^{{\dagger}(-)}$ Agr.~\cite{Drei-Pol-Thor}\\
\hline
$132$ $232$ & 76   & $[ {\superpartner{u}}_{L3} ]^{2}$ & ${\eta} {\to} {\mu} {\bar{e}} + e {\bar{{\mu}}}$ & & none & n/a & New\\
\cline{2-4}
            & 1.1  & ${\times} 10^{+5}$ $[ {\superpartner{u}}_{L3} ]^{2}$ & ${{\eta}'} {\to} {\mu} {\bar{e}}$ & & & & \\
\hline
$132$ $233$ & 9.4  & ${\times} 10^{-3}$ $[ {\superpartner{u}}_{L3} ]^{2}$ & $B_{s}^{0} {\to} e {\bar{{\mu}}}$ & 2.7 & ${\times} 10^{-4}$ $[ {\superpartner{u}}_{L3} ]^{2}$ & $b {\to} s e {\bar{{\mu}}}$ & Unimp.~\cite{Jang:1997ry}\\
\hline
$132$ $331$ & 2.3  & ${\times} 10^{-3}$ $[ {\superpartner{u}}_{L3} ]^{2}$ & ${\tau} {\to} e K_{S}$ & 2.7 & ${\times} 10^{-3}$ $[ {\superpartner{u}}_{L3} ]^{2}$ & ${\tau} {\to} e K^{0{\ast}}$ & Upd.~\cite{RPV_review382}\\
\cline{2-4}
            & 2.9  & ${\times} 10^{-3}$ $[ {\superpartner{u}}_{L3} ]^{2}$ & ${\tau} {\to} e K^{{\ast}0}$ & & & & \\
\hline
\end{tabular}
\end{center}
\caption{Bounds on $\mathbf{({{\lambda}'}_{ijk} {{\lambda}'}_{lmn})}$ continued.}
\label{lambdaprime-lambdaprime_table5}
\end{table}

\begin{table}[p]
\begin{center}
\begin{tabular}{|c||d@{${\columnspacefixer}$}r|l||d@{${\columnspacefixer}$}r|l|r|}
\hline
$( {{\lambda}'}_{ijk}$ ${{\lambda}'}_{lmn} )$ & \multicolumn{3}{c||}{From this work} & \multicolumn{4}{c|}{Previously published}\\
\cline{2-8}
$ijk$ $lmn$ & \multicolumn{2}{c|}{Bound}                                 & Decay                               & \multicolumn{2}{c|}{Bound}                     & Decay & Key\\
\hline
$221$ $223$ & 5.4  & ${\times} 10^{-4}$ $[ {\superpartner{u}}_{L2} ]^{2}$ & $B_{d}^{0} {\to} {\mu} {\bar{{\mu}}}$ & 2.1 & ${\times} 10^{-3}$ $[ {\superpartner{u}}_{L2} ]^{2}$ & $B_{d}^{0} {\to} {\mu} {\bar{{\mu}}}$ & Upd.~\cite{Saha_Kundu}\\
\hline
$221$ $321$ & 2.9  & $[ {\superpartner{d}}_{R1} ]^{2}$ & $J/{\psi} {\to} {\tau} {\bar{{\mu}}} / {\mu} {\bar{{\tau}}}$ & 0.21          & $[ {\superpartner{d}}_{R1} ]$ & ${\frac{{D^{0} {\to} {\nu} {\bar{{\mu}}} K^{-}}}{{D^{0} {\to} {\nu} {\bar{e}} K^{-}}}}$ & Unimp.~\cite{RPV_review277},\\
            &      &                                   &                                                              & {\times} 0.52 & $[ {\superpartner{d}}_{R1} ]$ & ${\frac{{D_{s}^{-} {\to} {\tau} {\bar{{\nu}}}}}{{D_{s}^{-} {\to} {\mu} {\bar{{\nu}}}}}}$ & \cite{RPV_review271}\\
\hline
$221$ $321$ & 1.6  & ${\times} 10^{-3}$ $[ {\superpartner{u}}_{L2} ]^{2}$ & ${\tau} {\to} {\mu} {\eta}$ & 4.4 & ${\times} 10^{-3}$ $[ {\superpartner{u}}_{L2} ]^{2}$ & ${\tau} {\to} {\mu} {\rho}^{0}$ & ${}^{{\dagger}(-)}$ Upd.~\cite{RPV_review382}\\
\cline{2-4}
            & 1.8  & ${\times} 10^{-3}$ $[ {\superpartner{u}}_{L2} ]^{2}$ & ${\tau} {\to} {\mu} {\pi}^{0}$ & & & & \\
\cline{2-4}
            & 4.3  & ${\times} 10^{-3}$ $[ {\superpartner{u}}_{L2} ]^{2}$ & ${\tau} {\to} {\mu} {\rho}^{0}$ & & & & \\
\hline
\end{tabular}
\end{center}
\caption{Bounds on $\mathbf{({{\lambda}'}_{ijk} {{\lambda}'}_{lmn})}$ continued.}
\label{lambdaprime-lambdaprime_table6}
\end{table}

\clearpage

\appendix

\section{Meson decay constants}
\label{app:qcd}

We have defined the decay constants of vector and pseudoscalar mesons
through
\begin{eqnarray}
{\bra{0}} {\bar{q}}_{{\alpha}} {\gamma}^{{\mu}} q_{{\beta}} {\ket{{V(
      p_V )}}} & \equiv & H_{V}^{{\alpha}{\beta}} f_{V} m_{V}
{\epsilon}_{V}^{{\mu}}\\
\mbox{and}\quad {\bra{0}} 
{\bar{q}}_{{\alpha}} {\gamma}^{{\mu}} {\gamma}^{5}
q_{{\beta}} {\ket{{P( p_P )}}} & \equiv & i H_{P}^{{\alpha}{\beta}} 
f_{P} p_{P}^{{\mu}}\,,
\label{PCAC_condition_equation}
\end{eqnarray}
where $H_{V/P}^{{\alpha}{\beta}}$ is the coefficient of
${\bar{q}}_{{\alpha}} q_{{\beta}}$ in the quark model wavefunction of
the meson. As $H_{V/P}^{{\alpha}{\beta}}$ is not standard notation we
shall describe it in some detail.  Firstly, it is only of relevance to
the light mesons composed of $u,d,s$-quarks, as it is assumed that the
charmed and bottom meson wavefunctions consist entirely of one quark
bilinear, \eg~$D^{0}$ is entirely ${\bar{d}} c$, so $H_{{D^{0}}}^{dc}
= 1$ and all other $H_{{D^{0}}}^{{\alpha}{\beta}} = 0$.  Hence for
mesons which are not part of the light SU(3)$_{uds}$ octet or singlet,
$H_{V/P}^{{\alpha}{\beta}} = 1$ for the relevant ${\alpha}$ and
${\beta}$. Similarly for the charged light mesons, \eg\ $K^{+}$ is
entirely ${\bar{s}} u$, hence $H_{{K^{+}}}^{su} = 1$.  For the neutral
light mesons, we obtain $H_{P}^{{\alpha}{\beta}}$ from the standard
PDG~\cite{PDG} definition of the pseudoscalar decay constant
\begin{equation}
{\sqrt{2}} {\bra{0}} {\bar{q}} {\gamma}^{{\mu}} {\gamma}^{5} 
{\frac{{\lambda}^{a}}{2}} q {\ket{{P^b( p )}}} = 
i \delta^{ab} f_{P} p^{{\mu}}\,,
\label{pseudoscalar_meson_decay_constant_definition_equation}
\end{equation}
where $q$ is the vector $q = ( u, d, s )^{T}$, and $a,b$ are
SU(3)-flavour indices. $P^{b}(p)=\bar{q}\lambda^b q$ denotes a basis
vector of the eight-dimensional representation of flavour SU(3), and
${\lambda}^{a}$ are the Gell-Mann matrices (normalized such that
${\trace{{{\lambda}^{a}{\lambda}^{b}}}} = 2 {\delta}^{ab}$; also here
${\lambda}^{0}$ is defined as ${\prettyfraction{1}{{\sqrt{3}}}}$ times
the three--by--three identity matrix). To relate
(\ref{PCAC_condition_equation}) and
(\ref{pseudoscalar_meson_decay_constant_definition_equation}) we note
that the quark bilinears $\bar{q}_\alpha q_\beta$ can be written as
linear combinations of $\bar{q} \lambda^a q$, so that
\begin{equation}
{\bra{0}} {\bar{q}}_{{\alpha}} {\gamma}^{{\mu}} {\gamma}^{5}
q_{{\beta}} {\ket{{P^b(p)}}}
= 
\sum_a C^a_{\alpha\beta}{\bra{0}} \bar{q} {\gamma}^{{\mu}} {\gamma}^{5}
\frac{\lambda^a}{2} q {\ket{{P^b(p)}}} = 
C^b_{\alpha\beta} \frac{i}{\sqrt{2}} f_{P} p_{{\mu}}\,.
\end{equation}
Expressing the physical meson states $\ket{P}$ in terms of the basis
states $\ket{P^b}$, we arrive at the generic equation
(\ref{PCAC_condition_equation}), where the coefficients
$H_{P^a}^{{\alpha}{\beta}}$ are given as
${\prettyfraction{1}{{\sqrt{2}}}} C_{\alpha\beta}^a$.

Let us consider a specific example and determine $\bra{0} \bar{u}
{\gamma}^{{\mu}} {\gamma}^{5} u \ket{\pi^0}$. We find 
\begin{equation}
\bra{0} \bar{u} {\gamma}^{{\mu}} {\gamma}^{5} u \ket{\pi^0(p)}
= 
\bra{0} \bar{q}\left(\sqrt{\frac{2}{3}}\frac{\lambda^0}{2} +
  \frac{\lambda^3}{2} + \frac{1}{\sqrt{3}} \frac{\lambda^8}{2} \right)
q \ket{\pi^3(p)}
= \frac{i}{\sqrt{2}} f_{\pi} p_{{\mu}}
\end{equation}
and hence $H_{{{\pi}^{0}}}^{uu} = {\prettyfraction{1}{{\sqrt{2}}}}$.
Note that with our definition
(\ref{pseudoscalar_meson_decay_constant_definition_equation})
$f_{{\pi}} = 130$~MeV.

In our numerical analysis we take into account
${\eta}^{0}$-${\eta}^{8}$ mixing, so ${\eta}$ and ${{\eta}'}$ are not
exactly ${\prettyfraction{1}{{\sqrt{6}}}} ( {\bar{u}} u + {\bar{d}} d
- 2 {\bar{s}} s )$ and ${\prettyfraction{1}{{\sqrt{3}}}} ( {\bar{u}} u
+ {\bar{d}} d + {\bar{s}} s )$, but mixtures with a mixing angle
${\theta}_{\eta} = -11.5^{{\circ}} = 0.052$ radians~\cite{PDG}, \eg\ 
${\bra{0}} {\bar{s}} {\gamma}^{5} {\gamma}^{{\mu}} s {\ket{{\eta}( p
    )}} = i [ {\cos}( {\theta}_{\eta} ) H_{{{\eta}_{8}}}^{ss}
f_{{{\eta}_{8}}} - {\sin}( {\theta}_{\eta} ) H_{{{\eta}_{0}}}^{ss}
f_{{{\eta}_{0}}} ] p^{{\mu}}$.  For ${\phi}$ and ${\omega}$ we assume
ideal mixing, so that $\phi = \bar{s}s$ and $\omega =
(\bar{u}u+\bar{d}d)/\sqrt{2}$. The non-trivial coefficients
$H_{P}^{{\alpha}{\beta}}$ can be read off the quark bilinear
coefficients listed in \citetable{quark_bilinear_coefficients_table}.
The $H_{V}^{{\alpha}{\beta}}$ are defined to be the same as the
$H_{P}^{{\alpha}{\beta}}$ for their pseudoscalar counterparts.

\begin{table}[t]
\begin{center}
\begin{tabular}{|l|l|}
\hline
${\pi}^{0}$ & ${\frac{1}{{\sqrt{2}}}} ( {\bar{u}} u - {\bar{d}} d )$\\
\hline
$K_{S}$     & ${\frac{1}{{\sqrt{2}}}} ( {\bar{s}} d + {\bar{d}} s )$\\
\hline
$K_{L}$     & ${\frac{1}{{\sqrt{2}}}} ( {\bar{s}} d - {\bar{d}} s )$\\
\hline
${\eta}$    & $0.515 ( {\bar{u}} u + {\bar{d}} d ) - 0.685 {\bar{s}} s$\\
\hline
${{\eta}'}$ & $0.484 ( {\bar{u}} u + {\bar{d}} d ) + 0.729 {\bar{s}} s$\\
\hline
${\phi}$    & ${\bar{s}} s$\\
\hline
\end{tabular}
\end{center}
\caption{Non--trivial quark bilinear coefficients.}
\label{quark_bilinear_coefficients_table}
\end{table}

Let us now discuss the derivation of the pseudoscalar matrix element
from the axial vector matrix element (\ref{PCAC_condition_equation})
in its general form
\begin{equation}
{\sqrt{2}} {\bra{0}} A_{\mu}^{a}(x) {\ket{{P^{b}( p )}}} = 
i {\delta}^{ab} f_{P} p_{{\mu}} {\exp}( -i {\dotproduct{p}{x}} )\,,
\label{pseudoscalar_meson_decay_constant_definition_equation_with_exp}
\end{equation}
with $A_{{\mu}}^{a} = {\bar{q}} {\gamma}_{{\mu}} {\gamma}^{5}
{\frac{1}{2}} {\lambda}^{a} q$.  Applying ${\partial}^{{\mu}}$ to both
sides leads to
\begin{equation}
{\sqrt{2}} {\bra{0}} {\partial}^{{\mu}} A_{{\mu}}^{a} {\ket{{P^{b}( p )}}} = 
{\delta}^{ab} f_{P} m_P^{2} {\exp}( -i {\dotproduct{p}{x}} )\,.
\label{differentiated_pseudoscalar_meson_decay_constant_definition_equation}
\end{equation}
Now
\begin{equation}
{\partial}^{{\mu}} A_{\mu}^{a} = 
{\partial}^{{\mu}} \left( {\bar{q}} {\gamma}_{{\mu}} 
{\gamma}^{5} {\frac{1}{2}} {\lambda}^{a} q \right) = 
\left( {\bar{q}} {\overleftarrow{{\dslash}}} {\gamma}^{5} {\frac{1}{2}} {\lambda}^{a} q 
+ {\bar{q}} {\dslash} {\gamma}^{5} {\frac{1}{2}} {\lambda}^{a} q \right) 
= {\bar{q}} {\gamma}^{5} {\frac{i}{2}} \left\{ {\lambda}^{a}, M \right\} q 
\label{differentiated_axial_current_definition_manipulation}
\end{equation}
assuming that the quark fields satisfy the Dirac equation, and $M$
here is defined as
\begin{equation}
M = \left( \begin{array}{ccc}
m_{u} & 0     & 0\\
0     & m_{d} & 0\\
0     & 0     & m_{s} \end{array} \right)\,.
\label{quark_model_mass_matrix_definition_equation}
\end{equation}
Combining this result with
\citeequation{differentiated_pseudoscalar_meson_decay_constant_definition_equation}
at $x=0$ leads to
\begin{equation}
{\bra{0}} {\bar{q}} {\gamma}^{5} {\frac{1}{2}} \left\{ {\lambda}^{a}, M \right\} q 
{\ket{{P^{b}( p )}}} = {\frac{-i}{{\sqrt{2}}}} {\delta}^{ab} f_{P} m_P^{2}\,.
\label{pseudscalar_mass_with_decay_constant_equation}
\end{equation}

Since
\begin{equation}
{\bra{0}} {\bar{q}}_{{\alpha}} {\gamma}^{5} 
( m_{{q_{{\alpha}}}} + m_{{q_{{\beta}}}} ) q_{{\beta}} {\ket{{P^{b}( p )}}} 
 =  \sum_a C^a_{\alpha\beta}
{\bra{0}} {\bar{q}} {\gamma}^{5} \frac12\left\{\lambda^{a}, M \right\} q 
{\ket{{P^{b}( p )}}} 
 = C_{{\alpha}{\beta}}^{b}{\frac{-i}{{\sqrt{2}}}} f_{P} m_P^{2}\,,
\label{mu_P_definition_equation}
\end{equation}
where $C_{{\alpha}{\beta}}^{a}$ is defined such that
${\bar{q}}_{{\alpha}} q_{{\beta}} = C_{{\alpha}{\beta}}^{a} {\bar{q}}
\frac{{\lambda}^{a}}{2} q$,
we arrive at 
\begin{equation}
{\bra{0}} {\bar{q}}_{{\alpha}} {\gamma}^{5} q_{{\beta}} {\ket{{P^{b}( p )}}} 
 = \frac{C_{{\alpha}{\beta}}^{b}}{( m_{{q_{{\alpha}}}} + m_{{q_{{\beta}}}} )}
{\frac{-i}{{\sqrt{2}}}} f_{P} p^{2}\,.
\end{equation}
By comparison with
\citeequation{pseudoscalar_meson_approximation_equationI},
${\mu}_{P}^{{\alpha}{\beta}}$ is identified as  
\begin{equation}
{\mu}_{P}^{{\alpha}{\beta}} {\mydefinedby} 
{\frac{{-H_{P}^{{\alpha}{\beta}}\sqrt{2} 
( m_{{q_{{\alpha}}}} + m_{{q_{{\beta}}}} )}}{ C_{{\alpha}{\beta}}^{b}}}\,.
\label{mu_P_identification_equation}
\end{equation}

Take the neutral pion as an example:
\begin{equation}
2 m_u {\bra{0}} \bar{u} {\gamma}^{5}  u {\ket{\pi^0( p )}}
 =  {\bra{0}} {\bar{q}} {\gamma}^{5} \left\{\left(\sqrt{\frac{2}{3}}
\frac{\lambda^0}{2} + \frac{\lambda^3}{2} + \frac{1}{\sqrt{3}} \frac{\lambda^8}{2} 
\right), M \right\} q {\ket{{\pi^{3}( p )}}}
 = \frac{-i}{\sqrt{2}} f_{\pi} m_{\pi}^{2}\,,
\end{equation}
so that ${\mu}_{{{\pi}^{0}}}^{uu} = -2 m_{u}$ and analogously
${\mu}_{{{\pi}^{0}}}^{dd} = 2 m_{d}$. We note that this result is in 
disagreement with \cite{Black:2002wh}.

\section{Feynman graphs and matrix elements}
\label{app:me}
In this Appendix we present the Feynman graphs and matrix elements of
the various decays.

\subsection{Charged lepton decaying into two charged 
leptons and one charged anti-lepton}
This process proceeds through the exchange of a sneutrino
${\tilde{{\nu}}}^{g}$ in the $t$- and $u$-channel:\\[5mm]
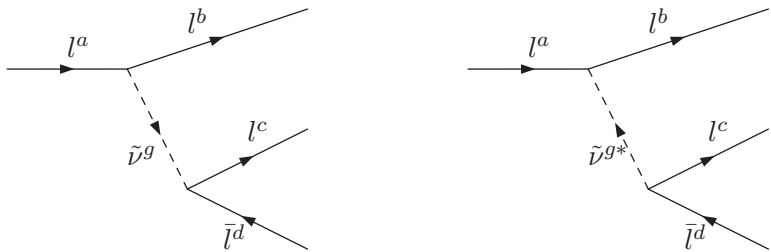
\hspace*{-1cm}
\begin{minipage}{6cm}
\begin{feynartspicture}( 256, 100 )( 1, 1 )
\FADiagram{} 
\FAProp( 0, 15 )( 10, 15 )( 0, ){/Straight}{1}
\FALabel( 5, 16 )[bl]{$l^{a}$} 
\FAProp( 10, 15 )( 25, 20 )( 0, ){/Straight}{1} 
\FALabel( 15, 18 )[bl]{$l^{b}$} 
\FAProp( 10, 15 )( 15, 5 )( 0, ){/ScalarDash}{1} 
\FALabel( 10, 7)[bl]{${\tilde{{\nu}}}^{g}$} 
\FAProp( 15, 5 )( 25, 10 )( 0, ){/Straight}{1} 
\FALabel( 20, 9 )[bl]{$l^{c}$} 
\FAProp( 15, 5 )( 25,0 )( 0, ){/Straight}{-1} 
\FALabel( 18, 0 )[bl]{${\bar{l}}^{d}$}
\end{feynartspicture}
\end{minipage}
\begin{minipage}{6cm}
\begin{feynartspicture}( 256, 100 )( 1, 1 )
\FADiagram{}
\FAProp( 0, 15 )( 10, 15 )( 0, ){/Straight}{1}
\FALabel( 5, 16 )[bl]{$l^{a}$}
\FAProp( 10, 15 )( 25, 20 )( 0, ){/Straight}{1}
\FALabel( 15, 18 )[bl]{$l^{b}$}
\FAProp( 10, 15 )( 15, 5 )( 0, ){/ScalarDash}{-1}
\FALabel( 10, 7 )[bl]{${\tilde{{\nu}}}^{g{\ast}}$}
\FAProp( 15, 5 )( 25, 10 )( 0, ){/Straight}{1}
\FALabel( 20, 9 )[bl]{$l^{c}$}
\FAProp( 15, 5 )( 25, 0 )( 0, ){/Straight}{-1}
\FALabel( 18, 0 )[bl]{${\bar{l}}^{d}$}
\end{feynartspicture}
\end{minipage}\\[2mm]

The matrix element for this decay is given by 
\begin{eqnarray}
i{\mathcal{M}}_{a{\to}bc{\bar{d}}} & = & {\langle} l^{b}( p_{{l^{b}}}), l^{c}
( p_{{l^{c}}} ), {\bar{l}}^{d}( p_{{{\bar{l}}^{d}}} ) | [ {\bar{l}}^{k} i 
{\lam}_{jik} P_{L} l^{i} {\tilde{{\nu}}}^{j} ] [ {\bar{l}}^{l} i 
{\lam}_{mln}^{{\ast}} P_{R} l^{n} {\tilde{{\nu}}}^{m{\ast}} ] | l^{a}
( p_{{l^{a}}} ) {\rangle} \nonumber\\ 
& = & 
{\frac{i}{{m_{{\tilde{{\nu}}}^{g}}^{2}}}} \bigg( [ {\bar{u}}( p_{{l^{b}}} ) 
{\lam}_{gba}^{{\ast}} P_{R} u( p_{{l^{a}}} ) ] [ {\bar{u}}( p_{{l^{c}}} ) 
{\lam}_{gdc} P_{L} v( p_{{{\bar{l}}^{d}}} ) ] \nonumber\\ & & \hspace*{6mm}
- [ {\bar{u}}( p_{{l^{b}}} ) {\lam}_{gdb} P_{L} v( p_{{{\bar{l}}^{d}}} ) ] 
[ {\bar{u}}( p_{{l^{c}}} ) {\lam}_{gca}^{{\ast}} P_{R} u( p_{{l^{a}}} ) ] 
\nonumber\\
 & & \hspace*{6mm} - [ {\bar{u}}( p_{{l^{c}}} ) {\lam}_{gac} P_{L}
u( p_{{l^{a}}} ) ] 
[ {\bar{u}}( p_{{l^{b}}} ) {\lam}_{gbd}^{{\ast}} P_{R} v( p_{{{\bar{l}}^{d}}} 
) ] \nonumber\\ & & \hspace*{6mm} 
+ [ {\bar{u}}( p_{{l^{b}}} ) {\lam}_{gab} P_{L} u( p_{{l^{a}}} ) ] [ 
{\bar{u}}( p_{{l^{c}}} ) {\lam}_{gcd}^{{\ast}} P_{R} v( p_{{{\bar{l}}^{d}}} ) 
] \bigg)
\label{M_a_to_bcd_equation}
\end{eqnarray}

\subsection{Charged lepton decays into a charged lepton 
and a neutral vector meson}
Charged leptons can decay into a vector meson and a charged lepton
through the exchange of a left-handed up-type squark or a right-handed
down-type squark:\\[5mm]
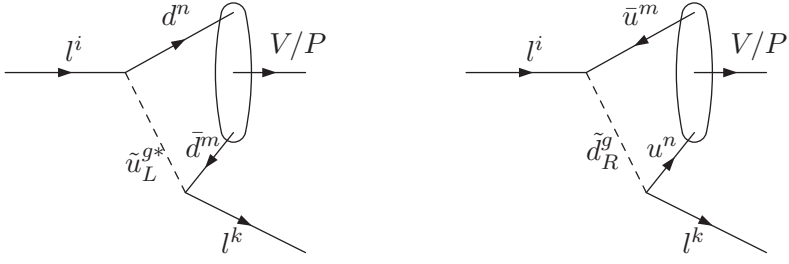
\hspace*{-1cm}
\begin{minipage}{6cm}
\begin{feynartspicture}( 256, 100 )( 1, 1 )
\FADiagram{}
\FAProp( 0, 15 )( 10, 15 )( 0, ){/Straight}{1}
\FALabel( 5, 16 )[bl]{$l^{i}$}
\FAProp( 10, 15 )( 19, 20 )( 0, ){/Straight}{1}
\FALabel( 13, 19 )[bl]{$d^{n}$}
\FAProp( 10, 15 )( 15, 5 )( 0, ){/ScalarDash}{0}
\FALabel( 10, 6 )[bl]{${\tilde{u}}_{L}^{g{\ast}}$}
\FAProp( 15, 5 )( 19, 10 )( 0, ){/Straight}{-1}
\FALabel( 15, 8 )[bl]{${\bar{d}}^{m}$}
\FAProp( 15, 5 )( 25, 0 )( 0, ){/Straight}{1}
\FALabel( 18, 0 )[bl]{$l^{k}$}
\FAProp( 18, 20 )( 20, 20 )( -0.8,){/Straight}{0}
\FAProp( 20, 20 )( 20, 10 )( -0.1,){/Straight}{0}
\FAProp( 20, 10 )( 18, 10 )( -0.8,){/Straight}{0}
\FAProp( 18, 10 )( 18, 20 )( -0.1,){/Straight}{0}
\FAProp( 19, 15 )( 25, 15 )( 0, ){/Straight}{1}
\FALabel( 22, 16 )[bl]{$V/P$}
\end{feynartspicture}
\end{minipage}
\begin{minipage}{6cm}
\begin{feynartspicture}( 256, 100 )( 1, 1 )
\FADiagram{}
\FAProp( 0, 15 )( 10, 15 )( 0, ){/Straight}{1}
\FALabel( 5, 16 )[bl]{$l^{i}$}
\FAProp( 10, 15 )( 19, 20 )( 0, ){/Straight}{-1}
\FALabel( 13, 19 )[bl]{${\bar{u}}^{m}$}
\FAProp( 10, 15 )( 15, 5 )( 0, ){/ScalarDash}{0}
\FALabel( 10, 7 )[bl]{${\tilde{d}}_{R}^{g}$}
\FAProp( 15, 5 )( 19, 10 )( 0, ){/Straight}{1}
\FALabel( 15, 8 )[bl]{$u^{n}$}
\FAProp( 15, 5 )( 25, 0 )( 0, ){/Straight}{1}
\FALabel( 18, 0 )[bl]{$l^{k}$}
\FAProp( 18, 20 )( 20, 20 )( -0.8,){/Straight}{0}
\FAProp( 20, 20 )( 20, 10 )( -0.1,){/Straight}{0}
\FAProp( 20, 10 )( 18, 10 )( -0.8,){/Straight}{0}
\FAProp( 18, 10 )( 18, 20 )( -0.1,){/Straight}{0}
\FAProp( 19, 15 )( 25, 15 )( 0, ){/Straight}{1}
\FALabel( 22, 16 )[bl]{$V/P$}
\end{feynartspicture}
\end{minipage}\\[2mm]

The matrix element for this process is given by
\begin{multline}
i{\mathcal{M}}_{l^i{\to}l^k+V}  = \\
 {\langle} {\mbox{out states}} | \big ( [ -i {{\lam}'}_
{ijn} {\bar{d}}_{R}^{n} l_{L}^{i} {\tilde{u}}_{L}^{j} ] [ -i {{\lam}'}_{ktm}^
{{\ast}}  {\bar{l}}_{L}^{k} d_{R}^{m} {\tilde{u}}_{L}^{t{\ast}} ] 
+ [ -i {{\lam}'}_{imj} l_{L}^{i} u_{L}^{m} {\tilde{d}}_{R}^{j} ] [ -i 
{{\lam}'}_{knt}^{{\ast}} {\bar{u}}_{L}^{n} {\bar{l}}_{L}^{k} {\tilde{d}}_{R}^
{t{\ast}} ] \big) | {\mbox{in states}} {\rangle}
\label{M_to_V_before_Fierz_equation}
\end{multline}
After some use of Fierz identities (and dropping the left/right
subscripts on the squarks, as only ``left--handed'' up--type squarks
or ``right--handed'' down--type squarks appear) we find
\begin{eqnarray}
i{\mathcal{M}}_{l^i{\to}l^k+V} & = & {\langle} {\mbox{out states}} | {\frac{1}{4}} 
[ {{\lam}'}_{ijn} {{\lam}'}_{ktm}^{{\ast}} {\tilde{u}}^{j} {\tilde{u}}^{t{\ast}} 
{\bar{d}}^{n} ( {\gamma}^{{\mu}} + {\gamma}^{{\mu}} {\gamma}^{5} ) d^{m} 
\nonumber\\ & & 
- {{\lam}'}_{imj} {{\lam}'}_{knt}^{{\ast}} {\tilde{d}}^{j} {\tilde{d}}^
{t{\ast}} {\bar{u}}^{n} ( {\gamma}^{{\mu}} - {\gamma}^{{\mu}} {\gamma}^{5} ) u^{m} ] 
[ {\bar{l}}^{k} P_{R} {\gamma}_{{\mu}} l^{i} ] | {\mbox{in states}} {\rangle}
\label{M_to_V_after_Fierz_equation}
\end{eqnarray}
Contracting the meson state with the quark bilinear results in 
\begin{eqnarray}
i{\mathcal{M}}_{l^i{\to}l^k+V} & = & 
{\langle}{\mbox{leptons out}} | 
{\frac{1}{4}} \left[ {\sum_{{\mbox{\scriptsize d--type}}}} {{\lam}'}_{ijn} 
{{\lam}'}
_{ktm}^{{\ast}} H_{V}^{mn{\ast}} {\tilde{u}}^{j} {\tilde{u}}^{t{\ast}} \right. 
\nonumber\\ & & 
- \left. {\sum_{{\mbox{\scriptsize u--type}}}} {{\lam}'}_{imj} {{\lam}'}_{knt}^
{{\ast}} 
H_{V}^{mn{\ast}} {\tilde{d}}^{j} {\tilde{d}}^{t{\ast}} \right] [ {\bar{l}}^{k} P_{R} 
{\gamma}_{{\mu}} l^{i} ] | {\mbox{leptons in}} {\rangle} f_{V}^{{\ast}} m_{V} 
{\epsilon}_{V}^{{\mu}{\ast}} \nonumber\\
 & = & {\frac{1}{4}} \left[ {\sum_{{\mbox{\scriptsize d--type}}}} {{\lam}'}_{ign} 
{{\lam}'}_{kgm}^{{\ast}} H_{V}^{mn{\ast}} \left( {\frac{-i}{{m_{{\tilde{u}}^{g}}^{2}}}}
\right) - {\sum_{{\mbox{\scriptsize u--type}}}} {{\lam}'}_{img}
 {{\lam}'}_{kng}^{{\ast}} H_{V}^{mn{\ast}} \left( {\frac{-i}{{
 m_{{\tilde{d}}^{g}}^{2}}}} \right) \right] \nonumber \\[2mm] 
&& \times [ {\bar{u}}( p_{{l^{k}}} ) P_{R} {\gamma}^{{\mu}} u( p_{{l^{i}}} ) ] 
f_{V}^{{\ast}} m_{V} {\epsilon}_{V{\mu}}^{{\ast}}
\label{M_to_V_final_equation}
\end{eqnarray}
where we have introduced the notation ${\sum_{{\mbox{\scriptsize
d--type}}}}$ to mean only summing over the down--type quarks in the 
meson and ${\sum_ {{\mbox{\scriptsize u--type}}}}$ to mean summing
over the up--type quarks. The $m$ and $n$ in $H_{V}^{mn}$ are the
generation indices of the appropriate quarks and can be used in
standard summation convention with the $m$ and $n$ appearing in the
coupling indices. For example, for the decay ${\tau} {\rightarrow} e
{\rho}^{0}$, we have
\begin{multline}
i{\mathcal{M}}_{{\tau} {\rightarrow} e {\rho}^{0}} =\\ {\frac{1}{4}} 
\left[ {\sum_{{\mbox{\scriptsize d--type}}}} {{\lam}'}_{3gn} {{\lam}'}_{1gm}^
{{\ast}} 
H_{{{\rho}^{0}}}^{mn{\ast}} \left( {\frac{-i}{{m_{{\tilde{u}}^{g}}^{2}}}} \right)
  -  {\sum_{{\mbox{\scriptsize u--type}}}} {{\lam}'}_{3mg} {{\lam}'}_{1ng}^
{{\ast}} 
H_{{{\rho}^{0}}}^{mn{\ast}} \left( {\frac{-i}{{m_{{\tilde{d}}^{g}}^{2}}}} \right)
 \right] [ {\bar{u}} ( p_{e} ) P_{R} {\gamma}^{{\mu}} u( p_{{\tau}} ) ]
 f_{{{\rho}^{0}}}^{{\ast}} m_{{{\rho}^{0}}} {\epsilon}_{{{\rho}^{0}}{\mu}}^{{\ast}}\\
 =  {\frac{-1}{4}} \left[ ( {{\lam}'}_{3g1} {{\lam}'}_{1g1}^{{\ast}}
 H_{{{\rho}^{0}}}^{dd{\ast}} + {{\lam}'}_{3g1} {{\lam}'}_{1g2}^{{\ast}} 
H_{{{\rho}^{0}}}^{ds{\ast}} + {{\lam}'}_{3g1} {{\lam}'}_{1g3}^{{\ast}} 
H_{{{\rho}^{0}}}^{db{\ast}} + {{\lam}'}_{3g2} {{\lam}'}_{1g1}^{{\ast}} 
H_{{{\rho}^{0}}}^{sd{\ast}} + {\mathellipsis} ) {\frac{i}{{m_{{\tilde{u}}^{g}}^{2}}}} 
\right. \\
  - \left. ( {{\lam}'}_{31g} {{\lam}'}_{11g}^{{\ast}} H_{{{\rho}^{0}}}^
{uu{\ast}} + {{\lam}'}_{31g} {{\lam}'}_{12g}^{{\ast}} H_{{{\rho}^{0}}}^
{uc{\ast}} + {\mathellipsis} ) {\frac{i}{{m_{{\tilde{d}}^{g}}^{2}}}} \right] 
[ {\bar{u}}( p_{e} ) P_{R} {\gamma}^{{\mu}} u( p_{{\tau}} ) ] f_{{{\rho}^{0}}}^
{{\ast}} m_{{{\rho}^{0}}} {\epsilon}_{{{\rho}^{0}}{\mu}}^{{\ast}} \\
  =  {\frac{-1}{4}} \left[ ( {{\lam}'}_{3g1} {{\lam}'}_{1g1}^
{{\ast}} * {\frac{-1}{{\sqrt{2}}}} + {{\lam}'}_{3g1} {{\lam}'}_{1g2}^
{{\ast}} * 0 + {{\lam}'}_{3g1} {{\lam}'}_{1g3}^{{\ast}} * 0 + {{\lam}'}_
{3g2} {{\lam}'}_{1g1}^{{\ast}} * 0 + {\mathellipsis} ) {\frac{i}{{m_{{\tilde{u}}^
{g}}^{2}}}} \right. \\
  - \left. ( {{\lam}'}_{31g} {{\lam}'}_{11g}^{{\ast}} * {\frac{1}
{{\sqrt{2}}}} + {{\lam}'}_{31g} {{\lam}'}_{12g}^{{\ast}} * 0 + {\mathellipsis} ) 
{\frac{i}{{m_{{\tilde{d}}^{g}}^{2}}}} \right] [ {\bar{u}}( p_{e} ) P_{R} {\gamma}^
{{\mu}} u( p_{{\tau}} ) ] f_{{{\rho}^{0}}}^{{\ast}} m_{{{\rho}^{0}}}
 {\epsilon}_{{{\rho}^{0}}{\mu}}^{{\ast}}
\label{tau_to_e_rho_example_M_equation}
\end{multline}

\subsection{Charged lepton decays into a charged lepton 
and a neutral pseudoscalar meson}

In addition to the two diagrams above, which can lead to pseudoscalar
mesons as well as vector mesons, there is a further diagram for the
decay into pseudoscalar mesons that is mediated by a sneutrino:

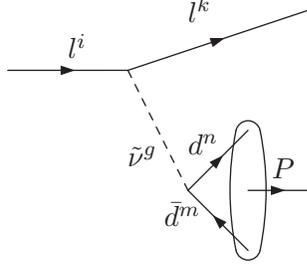
\begin{feynartspicture}( 256, 100 )( 1, 1 )

\FADiagram{}
\FAProp( 0, 15 )( 10, 15 )( 0, ){/Straight}{1}
\FALabel( 5, 16 )[bl]{$l^{i}$}
\FAProp( 10, 15 )( 25, 20 )( 0, ){/Straight}{1}
\FALabel( 15, 19 )[bl]{$l^{k}$}
\FAProp( 10, 15 )( 15, 5 )( 0, ){/ScalarDash}{0}
\FALabel( 10, 7 )[bl]{${\tilde{{\nu}}}^{g}$}
\FAProp( 15, 5 )( 20, 10 )( 0, ){/Straight}{1}
\FALabel( 15, 8 )[bl]{$d^{n}$}
\FAProp( 15, 5 )( 20, 0 )( 0, ){/Straight}{-1}
\FALabel( 13, 2 )[bl]{${\bar{d}}^{m}$}
\FAProp( 19, 10 )( 21, 10 )( -0.8,){/Straight}{0}
\FAProp( 21, 10 )( 21, 0 )( -0.1,){/Straight}{0}
\FAProp( 21, 0 )( 19, 0 )( -0.8,){/Straight}{0}
\FAProp( 19, 0 )( 19, 10 )( -0.1,){/Straight}{0}
\FAProp( 20, 5 )( 25, 5 )( 0, ){/Straight}{1}
\FALabel( 22, 6 )[bl]{$P$}
\end{feynartspicture}

The contribution from the squark--mediated diagrams is given by
\begin{eqnarray}
i{\mathcal{M}}_{l^i{\to}l^k+P}^{{\tilde{q}}} & = & {\langle} {\mbox{out
    states}} | 
{\frac{1}{4}} [ {{\lam}'}_{ijn} {{\lam}'}_{ktm}^{{\ast}} 
{\tilde{u}}^{j} {\tilde{u}}^{t{\ast}} {\bar{d}}^{n} ( {\gamma}^{{\mu}}
+ {\gamma}^{{\mu}} {\gamma}_{5} ) d^{m} \nonumber\\
 & & - {{\lam}'}_{imj} {{\lam}'}_{knt}^{{\ast}} {\tilde{d}}^{j} 
{\tilde{d}}^{t{\ast}} {\bar{u}}^{n} ( {\gamma}^{{\mu}} 
- {\gamma}^{{\mu}} {\gamma}_{5} ) u^{m} ] [ {\bar{l}}^{k} P_{R} 
{\gamma}_{{\mu}} l^{i} ] | {\mbox{in states}} {\rangle}
\label{M_to_P_squ_before_contraction_equation}
\end{eqnarray}

Contracting the meson state with the quark bilinear one finds
\begin{eqnarray}
i{\mathcal{M}}_{l^i{\to}l^k+P}^{{\tilde{q}}} & = & {\langle} {\mbox{lepton
    out states}} |  
{\frac{1}{4}} \left[ {\sum_{{\mbox{\scriptsize d--type}}}} {{\lam}'}_{ijn} 
{{\lam}'}_{ktm}^{{\ast}} H_{P}^{mn{\ast}} {\tilde{u}}^{j}
{\tilde{u}}^{t{\ast}} \right. \nonumber\\
 & & - \left. {\sum_{{\mbox{\scriptsize u--type}}}} {{\lam}'}_{imj}
   {{\lam}'}_{knt}^{{\ast}} H_{P}^{mn{\ast}} {\tilde{d}}^{j} 
{\tilde{d}}^{t{\ast}} \right] [{\bar{l}}^{k} P_{R} 
{\gamma}_{{\mu}} l^{i} ] | {\mbox{lepton in states}} {\rangle}
f_{P}^{{\ast}} p_{P}^{{\mu}} \nonumber\\
 & = & {\frac{1}{4}} \left[ {\sum_{{\mbox{\scriptsize d--type}}}}
   {{\lam}'}_{ign} {{\lam}'}_{kgm}^{{\ast}} H_{P}^{mn{\ast}}
\left( {\frac{-i}{{m_{{\tilde{u}}^{g}}^{2}}}} \right) -
 {\sum_{{\mbox{\scriptsize u--type}}}} {{\lam}'}_{img}{{\lam}'}_{kng}^{{\ast}} 
H_{P}^{mn{\ast}} \left( {\frac{-i}{{m_{{\tilde{d}}^{g}}^{2}}}} \right) \right]
 \nonumber\\[2mm]
 & & {\times} [ {\bar{u}}( p_{{l^{k}}} ) P_{R} {\gamma}_{{\mu}} 
u( p_{{l^{i}}} ) ] f_{P}^{{\ast}} p_{P}^{{\mu}}
\label{M_to_P_squ_final_equation}
\end{eqnarray}

The contribution from the sneutrino--mediated diagrams is given by
\begin{multline}
i{\mathcal{M}}_{l^i{\to}l^k+P}^{{\tilde{{\nu}}}} = {\langle} {\mbox{out
    states}} | [ {\bar{l}}^{k} i ( {\lam}_{jik} P_{L}
{\tilde{{\nu}}}^{j} 
+ {\lam}_{jki}^{{\ast}} {\tilde{{\nu}}}^{j{\ast}} P_{R} ) l^{i} ]
\\  {\times} [ {\bar{d}}^{n} i ( {{\lam}'}_{tmn}
P_{L} {\tilde{{\nu}}}^{t} + {{\lam}'}_{tnm}^{{\ast}} P_{R}
{\tilde{{\nu}}}^{t{\ast}} ) d^{m} ] | {\mbox{in states}} {\rangle} \\ 
= {\frac{-i}{2 m_{{\tilde{{\nu}}}^{g}}^{2}}}
{\sum_{{\mbox{\scriptsize d--type}}}} [ {\bar{u}}( p_{{l^{k}}} ) {\lam}_{gik}
P_{L} u( p_{{l^{i}}} ) {{\lam}'}_{gnm}^{{\ast}} 
- {\bar{u}}( p_{{l^{k}}} ) {\lam}_{gki}^{{\ast}} P_{R} u(
 p_{{l^{i}}} ) {{\lam}'}_{gmn} ] \left[ {\frac{{H_{P}^{mn} 
f_{P} m_{P}^{2}}}{{{\mu}_{P}^{mn}}}} \right]^{{\ast}}
\label{M_to_P_snu_equation}
\end{multline}
noting that the sneutrino does not couple to up--type quarks.

For the case of a meson decaying into a lepton and an anti--lepton,
the matrix elements are identical up to making the appropriate index
substitutions in the couplings.

%
%


\begin{thebibliography}{99}

\bibitem{Glashow:1961tr}
  S.~L.~Glashow,
  Nucl.\ Phys.\  {\bf 22} (1961) 579.
\bibitem{Weinberg:1967tq}
  S.~Weinberg,
  Phys.\ Rev.\ Lett.\  {\bf 19} (1967) 1264.

\bibitem{Wess:1974tw}
  J.~Wess and B.~Zumino,
  Nucl.\ Phys.\ B {\bf 70} (1974) 39.

\bibitem{Sakai:1981pk}
  N.~Sakai and T.~Yanagida,
  Nucl.\ Phys.\ B {\bf 197} (1982) 533.

\bibitem{Weinberg:1981wj}
  S.~Weinberg,
  Phys.\ Rev.\ D {\bf 26} (1982) 287.

\bibitem{Shiozawa:1998si}
  M.~Shiozawa {\it et al.}  [Super-Kamiokande Collaboration],
  Phys.\ Rev.\ Lett.\  {\bf 81} (1998) 3319
  [arXiv:hep-ex/9806014].

\bibitem{Dreiner:2005rd}
  H.~K.~Dreiner, C.~Luhn and M.~Thormeier,
  Phys.\ Rev.\ D {\bf 73} (2006) 075007
  [arXiv:hep-ph/0512163].

\bibitem{Dimopoulos:1981dw}
  S.~Dimopoulos, S.~Raby and F.~Wilczek,
  Phys.\ Lett.\ B {\bf 112} (1982) 133.

\bibitem{MSSM_review}
  For an introduction, see \eg:
  S.~P.~Martin,
  arXiv:hep-ph/9709356;
  I.~J.~R.~Aitchison,
  arXiv:hep-ph/0505105;
M.~Drees,
  arXiv:hep-ph/9611409.

\bibitem{Ibanez:1991pr}
  L.~E.~Ibanez and G.~G.~Ross,
  Nucl.\ Phys.\ B {\bf 368} (1992) 3.


\bibitem{Dreiner:2006xw}
  H.~K.~Dreiner, C.~Luhn, H.~Murayama and M.~Thormeier,
  arXiv:hep-ph/0610026.

\bibitem{discrete-gauge}
L.~M.~Krauss and F.~Wilczek,
  Phys.\ Rev.\ Lett.\  {\bf 62} (1989) 1221;
L.~E.~Ibanez and G.~G.~Ross,
  Phys.\ Lett.\ B {\bf 260} (1991) 291.

\bibitem{Hall:1983id}
  L.~J.~Hall and M.~Suzuki,
  Nucl.\ Phys.\ B {\bf 231} (1984) 419.

\bibitem{Nardi:1996iy}
  E.~Nardi,
  Phys.\ Rev.\ D {\bf 55} (1997) 5772
  [arXiv:hep-ph/9610540].

\bibitem{Allanach:2003eb}
  B.~C.~Allanach, A.~Dedes and H.~K.~Dreiner,
  Phys.\ Rev.\ D {\bf 69} (2004) 115002
  [Erratum-ibid.\ D {\bf 72} (2005) 079902]
  [arXiv:hep-ph/0309196].


\bibitem{Hirsch:2000ef}
  M.~Hirsch, M.~A.~Diaz, W.~Porod, J.~C.~Romao and J.~W.~F.~Valle,
  Phys.\ Rev.\ D {\bf 62} (2000) 113008
  [Erratum-ibid.\ D {\bf 65} (2002) 119901]
  [arXiv:hep-ph/0004115];
A.~Abada, G.~Bhattacharyya and M.~Losada,
  Phys.\ Rev.\ D {\bf 66} (2002) 071701
  [arXiv:hep-ph/0208009];
E.~J.~Chun, D.~W.~Jung and J.~D.~Park,
  Phys.\ Lett.\ B {\bf 557} (2003) 233
  [arXiv:hep-ph/0211310];
D.~E.~Kaplan and A.~E.~Nelson,
  JHEP {\bf 0001} (2000) 033
  [arXiv:hep-ph/9901254];
R.~Hempfling,
  Nucl.\ Phys.\ B {\bf 478} (1996) 3
  [arXiv:hep-ph/9511288].


\bibitem{colliders}
S.~Dimopoulos, R.~Esmailzadeh, L.~J.~Hall and G.~D.~Starkman,
  Phys.\ Rev.\ D {\bf 41} (1990) 2099;
H.~K.~Dreiner and G.~G.~Ross,
  Nucl.\ Phys.\ B {\bf 365} (1991) 597.

\bibitem{PDG}
  W.~M.~Yao {\it et al.}  [Particle Data Group],
  J.\ Phys.\ G {\bf 33} (2006) 1.

\bibitem{Bhattacharyya:1995pr}
  G.~Bhattacharyya, J.~R.~Ellis and K.~Sridhar,
  Mod.\ Phys.\ Lett.\ A {\bf 10} (1995) 1583
  [arXiv:hep-ph/9503264].

\bibitem{Agashe:1995qm}
  K.~Agashe and M.~Graesser,
  Phys.\ Rev.\ D {\bf 54} (1996) 4445
  [arXiv:hep-ph/9510439].

\bibitem{Choudhury_Roy}
  D.~Choudhury and P.~Roy,
  Phys.\ Lett.\ B {\bf 378} (1996) 153
  [arXiv:hep-ph/9603363].

\bibitem{Jang_Kim_Lee}
  J.~H.~Jang, J.~K.~Kim and J.~S.~Lee,
  Phys.\ Rev.\ D {\bf 55} (1997) 7296
  [arXiv:hep-ph/9701283].

\bibitem{Kim_Ko_Lee}
  J.~E.~Kim, P.~Ko and D.~G.~Lee,
  Phys.\ Rev.\ D {\bf 56} (1997) 100
  [arXiv:hep-ph/9701381].

\bibitem{Bhattacharyya:1998be}
  G.~Bhattacharyya and A.~Raychaudhuri,
  Phys.\ Rev.\ D {\bf 57} (1998) 3837
  [arXiv:hep-ph/9712245].

\bibitem{Lebedev:1999vc}
  O.~Lebedev, W.~Loinaz and T.~Takeuchi,
  Phys.\ Rev.\ D {\bf 61} (2000) 115005
  [arXiv:hep-ph/9910435];
O.~Lebedev, W.~Loinaz and T.~Takeuchi,
  Phys.\ Rev.\ D {\bf 62} (2000) 015003
  [arXiv:hep-ph/9911479];
S.~A.~Abel, A.~Dedes and H.~K.~Dreiner,
  JHEP {\bf 0005} (2000) 013
  [arXiv:hep-ph/9912429].

\bibitem{Drei-Pol-Thor}
  H.~K.~Dreiner, G.~Polesello and M.~Thormeier,
  Phys.\ Rev.\ D {\bf 65} (2002) 115006
  [arXiv:hep-ph/0112228].

\bibitem{Saha_Kundu}
  J.~P.~Saha and A.~Kundu,
  Phys.\ Rev.\ D {\bf 66} (2002) 054021
  [arXiv:hep-ph/0205046].

\bibitem{Jang:1997ry}
  J.~H.~Jang, Y.~G.~Kim and J.~S.~Lee,
  Phys.\ Rev.\ D {\bf 58} (1998) 035006
  [arXiv:hep-ph/9711504].

\bibitem{Dreiner:2002xg}
  H.~Dreiner, G.~Polesello and M.~Thormeier,
  arXiv:hep-ph/0207160.

\bibitem{Xu_Wang_Yang}
  Y.~G.~Xu, R.~M.~Wang and Y.~D.~Yang,
  arXiv:hep-ph/0610338.

\bibitem{Barger_Giudice_Han}
  V.~D.~Barger, G.~F.~Giudice and T.~Han,
  Phys.\ Rev.\ D {\bf 40} (1989) 2987.

\bibitem{Davidson:1993qk}
  S.~Davidson, D.~C.~Bailey and B.~A.~Campbell,
  Z.\ Phys.\ C {\bf 61} (1994) 613
  [arXiv:hep-ph/9309310].

\bibitem{Dreiner:1997uz}
  H.~K.~Dreiner,
  arXiv:hep-ph/9707435.

\bibitem{Bhattacharyya:1997vv}
  G.~Bhattacharyya,
  arXiv:hep-ph/9709395.

\bibitem{Allanach:1999ic}
  B.~C.~Allanach, A.~Dedes and H.~K.~Dreiner,
  Phys.\ Rev.\ D {\bf 60} (1999) 075014
  [arXiv:hep-ph/9906209].

\bibitem{Herz}
  M.~Herz, Diploma Thesis,
  arXiv:hep-ph/0301079.

\bibitem{Chemtob:2004xr}
  M.~Chemtob,
  Prog.\ Part.\ Nucl.\ Phys.\  {\bf 54} (2005) 71
  [arXiv:hep-ph/0406029].

\bibitem{Barbier:2004ez}
  R.~Barbier {\it et al.},
  Phys.\ Rept.\  {\bf 420} (2005) 1
  [arXiv:hep-ph/0406039].
  
\bibitem{footnote1} Our effective Lagrangian is to be compared with
  those in \cite{Choudhury_Roy}, \cite{Jang_Kim_Lee},
  \cite{Kim_Ko_Lee}, \cite{Saha_Kundu} and \cite{Barger_Giudice_Han}
  (noting that both \cite{Jang_Kim_Lee} and \cite{Barger_Giudice_Han}
  use the convention where there is \textit{no} factor of
  ${\prettyfraction{1}{2}}$ before the ${\lam}$ terms).  We disagree
  with the form of the effective Lagrangians in \cite{Choudhury_Roy}
  and \cite{Saha_Kundu}, and agree with those in \cite{Jang_Kim_Lee},
  \cite{Kim_Ko_Lee} and \cite{Barger_Giudice_Han}. We note, however,
  that through projection onto vector or pseudoscalar quark bilinears
  the difference with \cite{Saha_Kundu} reduces to a simple overall
  sign error of the matrix element, which is then eliminated by
  squaring.  Accordingly, this has no ill effects in the quadratic
  coupling dominance convention.  Also, in the case of
  \cite{Choudhury_Roy} we note that it is merely that the wrong
  coupling in the second term of their equation (7) has the $*$
  denoting complex conjugation.
  
\bibitem{pcac} 
    See for example T.~P.~Cheng and L.~F.~Li, ``Gauge Theory Of
  Elementary Particle Physics,'' Oxford, Uk: Clarendon ( 1984) 536 P.
  ( Oxford Science Publications).
  
\bibitem{C_V_reference} M.~Neubert and B.~Stech,
  Adv.\ Ser.\ Direct.\ High Energy Phys.\  {\bf 15} (1998) 294
  [arXiv:hep-ph/9705292].

\bibitem{footnote2}
  In all these cases, it turns out we are agreeing with
  \cite{Drei-Pol-Thor}, though
  \cite{Drei-Pol-Thor} only gives the bounds to the
  first significant figure, and we assume that the differences ($6.7
  {\times} 10^{-9}$ compared to $6 {\times} 10^{-9}$ and $2.7 {\times}
  10^{-7}$ compared to $3 {\times} 10^{-7}$) arise from rounding
  errors.
  
\bibitem{footnote3} We do not agree with the bounds presented in
  \cite{Barbier:2004ez}, which are just those taken from
  \cite{Saha_Kundu}.  We believe that \cite{Saha_Kundu} presented
  incorrect bounds, and that this arises from their equation (13) for
  the decay width ${\Gamma}( B_{{q_{i}}} {\to} l_{l}^{-} l_{m}^{+} )$.
  We find that the ratio of the decay width (13) in \cite{Saha_Kundu}
  over our decay width (\ref{Gamma_P_to_snu_equation}) is ${4 ( m_{b}
    + m_{{q_{i}}} )^{2}}/{{M_{{B_{{q_{i}}}}}^{2}}}$.  While $m_{b} +
  m_{{q_{i}}}\approx M_{{B_{{q_{i}}}}}$ for heavy mesons, we believe
  that equation (13) in \cite{Saha_Kundu} misses a factor
  $\prettyfraction{1}{4}$ which results in too tight bounds. However,
  as the experimental bounds for many of the rare $B$ decay branching
  ratios have improved, we still obtain bounds for the couplings
  associated with these decays tighter than in \cite{Saha_Kundu}. Note
  that we agree with the corresponding equation~(8) in
  \cite{Jang_Kim_Lee} (taking into account that \cite{Jang_Kim_Lee}
  defines the couplings ${\lam}$ such that the superpotential does
  \textit{not} have the factor of ${\prettyfraction{1}{2}}$ before the
  trilinear lepton term) and with the generic result in
  \cite{Skiba:1992mg}.

\bibitem{Skiba:1992mg}
  W.~Skiba and J.~Kalinowski,
  Nucl.\ Phys.\ B {\bf 404} (1993) 3.

\bibitem{RPV_review382}
  \cite{Barbier:2004ez} updating bounds given in \cite{Kim_Ko_Lee} with
  data published by the Particle Data Group \cite{PDG_PhysRevD66} in
  2002.

\bibitem{PDG_PhysRevD66}
  K.~Hagiwara {\it et al.}  [Particle Data Group],
  Phys.\ Rev.\ D {\bf 66} (2002) 010001.

\bibitem{RPV_review359} 
  \cite{Barbier:2004ez} updating bounds given in \cite{Choudhury_Roy}
  with data published by the Particle Data Group \cite{PDG_PhysRevD66}
  in 2002

\bibitem{RPV_review382-383}
  \cite{Barbier:2004ez} updating bounds given in \cite{Kim_Ko_Lee} with
  data from SINDRUM II Collaboration, talk given at 14th
  International Conference on Particles and Nuclei (PANIC),
  Williamsburg, Virginia, USA, 22-28 May 1996. Presented by
  P. Wintz. Published by World Scientific Publishing
  Co. Pte. Ltd. 1997, Ed. C.E. Carlson and J.J. Domingo, 458.

\bibitem{RPV_review277}
  G.~Altarelli, G.~F.~Giudice and M.~L.~Mangano, 
  Nucl.\ Phys.\ B {\bf 506} (1997) 29 [arXiv:hep-ph/9705287].\\ 
  updated by \cite{Barbier:2004ez} with data published by the
  Particle Data Group \cite{PDG_PhysRevD66} in 2002.

\bibitem{RPV_review271}
  F.~Ledroit and G.~Sajot, Rapport GDR-Supersym\'{e}trie, GDR-S-008
  (ISN, Grenoble, 1998)\\ updated by \cite{Barbier:2004ez} with data
  published by the Particle Data Group \cite{PDG_PhysRevD66} in 2002.

\bibitem{Buras:2001pn}
  see \eg\ A.~J.~Buras,
  arXiv:hep-ph/0101336.

\bibitem{Black:2002wh}
  D.~Black, T.~Han, H.~J.~He and M.~Sher,
  Phys.\ Rev.\ D {\bf 66} (2002) 053002
  [arXiv:hep-ph/0206056].

\end{thebibliography}
\end{document}